\documentclass[traditabstract]{aa} 
\usepackage{amsmath}
\usepackage{graphicx}
\usepackage{txfonts}

\usepackage{natbib}
\bibpunct{(}{)}{;}{a}{}{,}
\usepackage{float}
\usepackage{calc}
\usepackage{textcomp}
\usepackage{placeins}
\usepackage{color}
\usepackage{rotating} 
\usepackage{lscape}
\usepackage{url}
\usepackage[colorlinks=true,linkcolor=blue,citecolor=blue]{hyperref}
\usepackage{subfig}
\usepackage[all]{draftcopy}
\usepackage{longtable}
\usepackage{array}
\usepackage{threeparttable}
\usepackage{lscape}
\DeclareTextSymbol{\degre}{OT1}{23}
\newcounter{savedfootnote}

\def \microns{{\,$\mu$m}}

\begin{document}
\title{On the SFR-M$_*$ main sequence archetypal star-formation history and analytical models.}

\author{L.~Ciesla\inst{1},  D.~Elbaz\inst{1}, and J.Fensch\inst{1}.
}

\institute{	
 Laboratoire AIM-Paris-Saclay, CEA/DSM/Irfu - CNRS - Universit\'e Paris Diderot, CEA-Saclay, F-91191 Gif-sur-Yvette, France
}	
 
   \date{Received; accepted}

  \abstract
{
The star formation history (SFH) of galaxies is a key assumption to derive their physical properties and can lead to strong biases.
In this work, we derive the SFH of main sequence (MS) galaxies showing how the peak SFH of a galaxy depends on its seed mass at e.g. $z$$=$5.
This seed mass reflects the galaxy's underlying dark matter (DM) halo environment.
We show that, following the MS, galaxies undergo a drastic slow down of their stellar mass growth after reaching the peak of their SFH.
According to abundance matching, these masses correspond to a hot and massive DM halos which state could results in less efficient gas inflows on the galaxies and thus could be at the origin of the limited stellar mass growth.
As a result, we show that galaxies, still on the MS, can enter the passive region of the UVJ diagram while still forming stars.
The best fit to the MS SFH is provided by a Right Skew Peak Function for which we provide parameters depending on the seed mass of the galaxy.
The ability of the classical analytical SFHs to retrieve the SFR of galaxies from Spectral Energy Distribution (SED) fitting is studied.
Due to mathematical limitations, the exponentially declining and delayed SFH struggle to model high SFR which starts to be problematic at $z$$>$2.
The exponentially rising and log-normal SFHs exhibit the opposite behavior with the ability to reach very high SFR, and thus model starburst galaxies, but not low values such as those expected at low redshift for massive galaxies.
Simulating galaxies SED from the MS SFH, we show that these four analytical forms recover the SFR of MS galaxies with an error dependent on the model and the redshift.
They are, however, sensitive enough to probe small variations of SFR within the MS, with an error ranging from 5 to 40$\%$ depending on the SFH assumption and redshift, but all the four fail to recover the SFR of rapidly quenched galaxies.
However, these SFHs lead to an artificial gradient of age, parallel to the MS which is not exhibited by the simulated sample.
This gradient is also produced on real data as we show using a sample of GOODS-South galaxies with redshifts between 1.5 and 2.5.
Here, we propose a SFH composed of a delayed form to model the bulk of stellar population with the addition of a flexibility in the recent SFH.
This SFH provides very good estimates of the SFR of MS, starbursts, and rapidly quenched galaxies at all redshift.
Furthermore, used on the GOODS-South sample, the age gradient disappears, showing its dependency on the SFH assumption made to perform the SED fitting.
}

  
   \authorrunning{Ciesla et al.}
   \titlerunning{Archetypal star-formation history and analytical models }
   
   \maketitle
   
\section{\label{intro}Introduction}
The evolution of galaxies depends on their star formation history (SFH) which is by definition the star formation rate of a galaxy, as a function of time.
Two main physical properties of galaxies are directly computed from their Spectral Energy Distributions (SED) assuming an SFH: the stellar mass and the star formation rate (SFR) at the time the galaxy is observed.
For star-forming galaxies, these two parameters follow a relation called the main sequence of galaxies \citep[MS, e.g.][]{Noeske07_SFseq,Elbaz07,Elbaz11,Rodighiero11,Schreiber15} whose normalization increases with redshift.
Its scatter however is found to be roughly constant over cosmic time \citep{Ilbert15,Schreiber15}.
The main consequence of this relation is that galaxies are forming the bulk of their stars through steady state processes rather than violent episodes of star formation, putting constraints on the SFH of galaxies.
One method to derive the stellar mass and SFR of galaxies is to build and model their spectral energy distribution.
To do so, one has to assume a stellar population model \citep[e.g.,][]{BruzualCharlot03,Maraston05} convolved by a star formation history, and then to apply an attenuation law \citep[][]{Calzetti00}.
The models built to fit the data are thus dependent on the SFH of galaxies.

First order approximation of galaxy SFHs can be guessed through the evolution of the SFR density of galaxies as a function of cosmic time \citep[see the review by][]{MadauDickinson14} showing that the global SFR of galaxies peaks around $z\sim2$ and then smoothly decreases.
Going further, several studies showed that sophisticated SFH parametrizations including stochastic events, such as those predicted by hydrodynamical simulations and semi-analytical models, have to be taken into account to reproduce galaxies SFH \citep[e.g.,][]{Lee10,Pacifici13,Behroozi13,Pacifici16}.
These models are complex to implement and a large library is needed to be able to model all galaxies properties.
Instead, numerous studies tested and used simple analytical forms to easily compute galaxies physical properties \citep[e.g.,][]{Papovich01,Maraston10, Pforr12,Gladders13,Simha14,Buat14,Boquien14,Ciesla15,Abramson16,Ciesla16}.
Recently, \cite{Ciesla15} used complex SFHs produced by the semi-analytical model GALFORM  \citep{Cole00,Bower06} to model $z$$=$1 galaxies Spectral Energy Distribution (SED) and recover their stellar mass and SFR using simple analytical SFHs.
They found that the two-exponentially decreasing SFH and the delayed SFH relatively well recovered the real SFR and M$_*$.
There is, however, a general agreement on the difficulty to constrain the age of the galaxy, here defined as the age of the oldest star, from broad-band SED fitting \citep[e.g.][]{Maraston10,Pforr12,Buat14,Ciesla15}.

The exquisite sensitivity of HST and \textit{Spitzer} now combined with ALMA observations, and soon to be complemented by JWST, allows us to better sample high redshift galaxy SEDs and thus apply more sophisticated SED fitting method implying the use of these analytical SFHs tested at lower redshifts.
The aim of this study is to compute the SFH of galaxies following the MS and understand if the widely used analytical SFHs are pertinent to model them, as well as outliers such as starburst or rapidly quenched galaxies, regardless of the redshift.

Throughout this paper, we use the WMAP7 cosmological parameters \citep[$H_0=70.4$\,km\,Mpc$^{-1}$\,s$^{-1}$, $\Omega_0=0.272$,][]{Komatsu11}.
SED fitting is performed assuming an IMF of \cite{Salpeter55}, but the results of this work are found to be robust against IMF choice.

\section{ \label{sfhms}The star formation history of a main sequence galaxy}
Over the past decade, numerous studies have shown that the bulk of star-forming galaxies follow a relation between their SFR and their stellar mass with observational confirmation that this relation holds up to $z$$=$4 \citep{Schreiber17}.
Recently, \cite{Schreiber15} parametrized the MS as a function of stellar mass and redshift.
It is thus possible to build the SFH of a star-forming galaxy following the MS by computing the SFR corresponding to the stellar mass at each redshift or time step.
Due to the decrease of the MS normalization with decreasing redshift, the typical SFR of galaxies at a given fixed mass (e.g. around M$^*$, the knee of the stellar mass function) declines with time.
However, because of the positive slope of the MS, the SFR of a galaxy increases with its stellar mass up to high masses where the MS starts to bend.
Computing the resulting SFH of a star-forming galaxy staying on the MS when forming stars will allow us to understand how the two effects affect the star formation activity of the galaxy.

\begin{figure}
  	\includegraphics[width=\columnwidth]{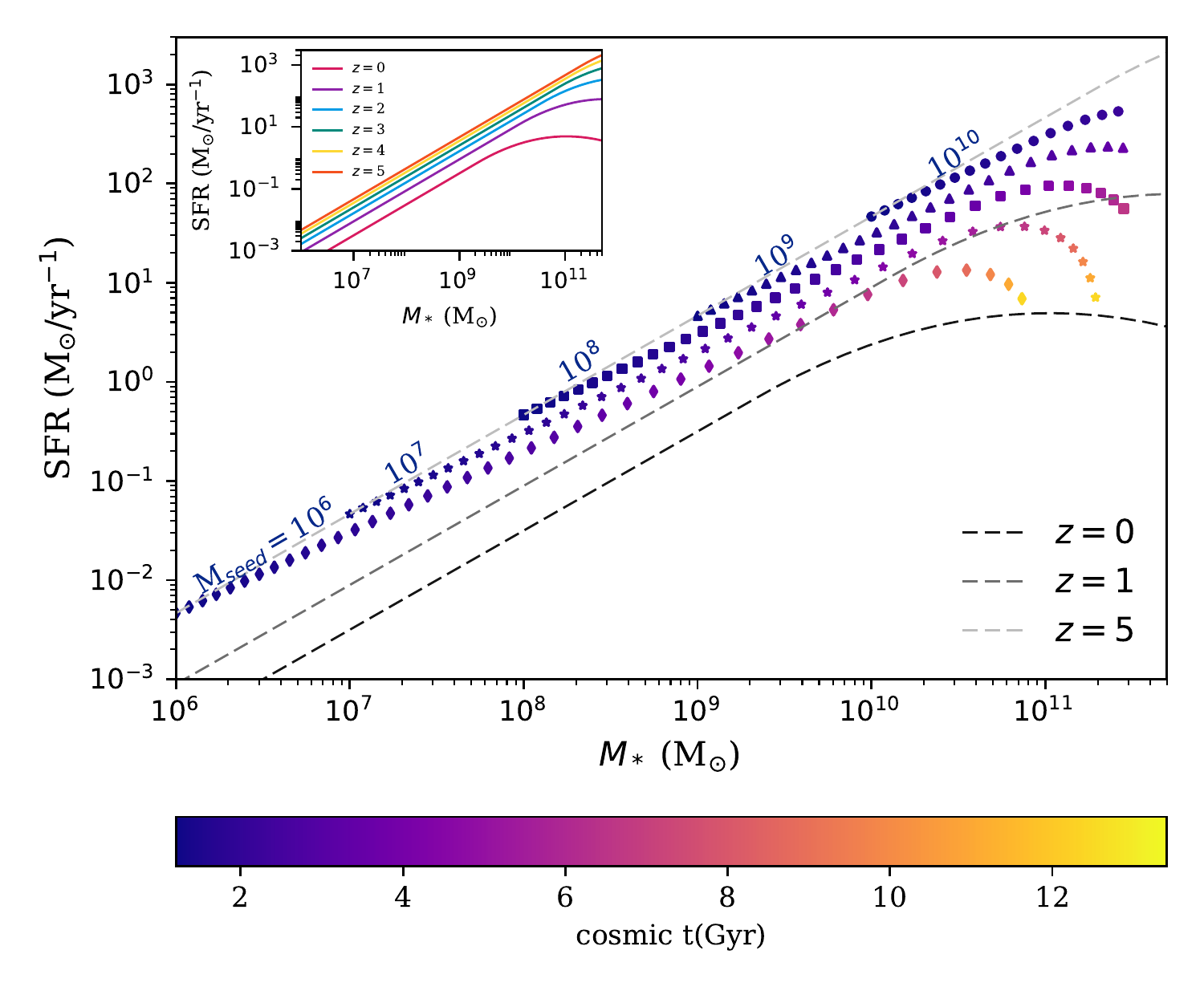}
  	\caption{ \label{evolution} Evolution of the simulated galaxies following the main sequence from $z$$=$5 to $z$$=$0.3 on the SFR-M$_*$ diagram, color coded with cosmic time. The different symbols correspond to different mass seeds: 10$^6$ (diamonds), 10$^7$ (stars), 10$^8$ (squares), 10$^9$ (triangles), and 10$^{10}$ M$_{\odot}$ (circles). The dashed lines indicates the MS at redshift 0 (black), 1 (gray), and 5 (light gray). The insert panel shows the MS relations from $z$$=$5 to $z$$=$0 on which the evolution of MS galaxies is computed.} 
\end{figure}

Starting at $z$$=$5 with a given mass seed, we calculate at each time step the SFR corresponding to the redshift and mass, assuming the MS relations of \cite{Schreiber15}.
Then the mass produced during the time step is calculated assuming a constant SFR.
At the following time step, we derive the new SFR associated with the redshift at the given time and the new computed mass.
The computed SFR and stellar masses are shown on the MS plane in Fig.~\ref{evolution} for every five time steps, for five seed mass.

\begin{figure}
  	\includegraphics[width=\columnwidth]{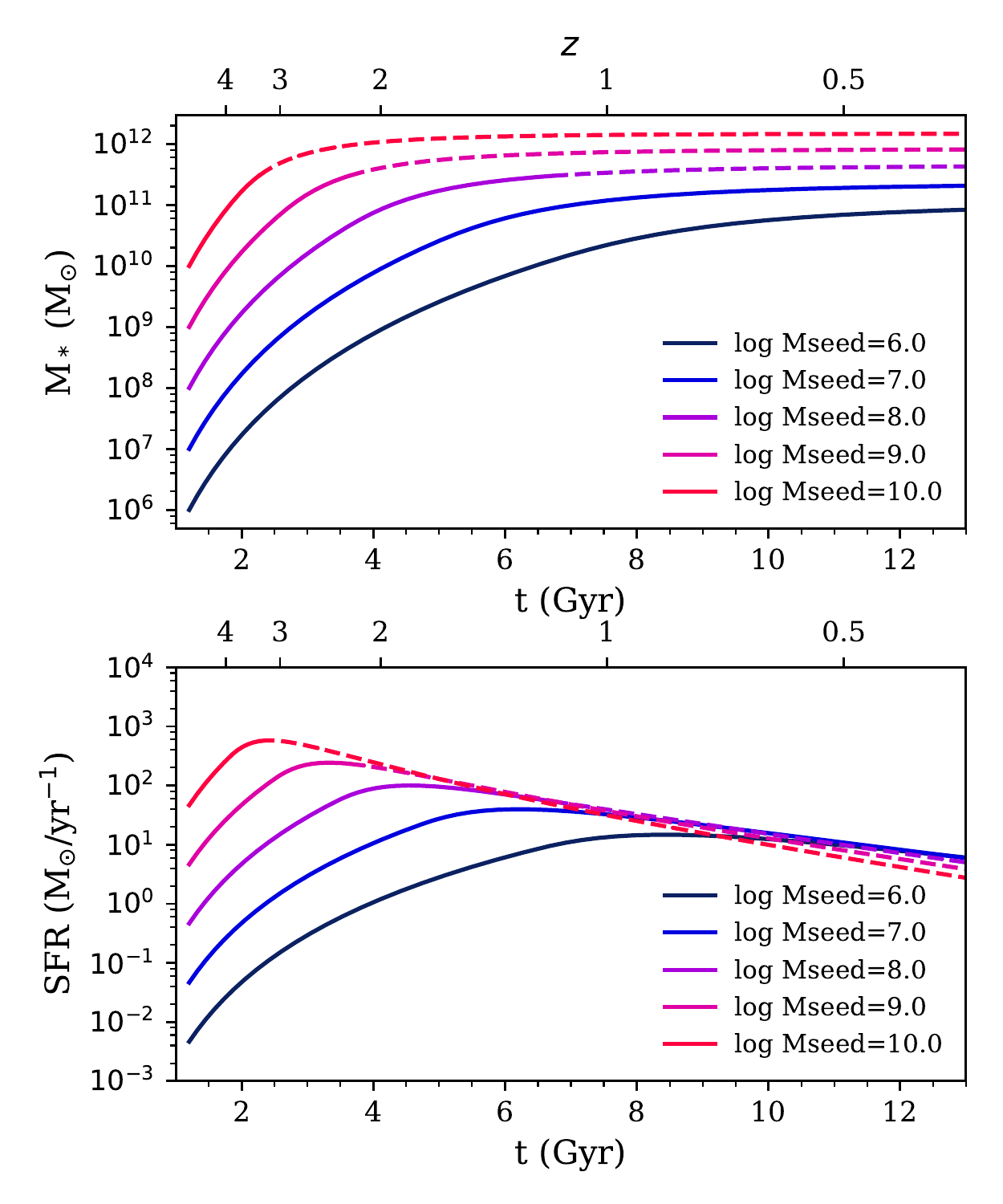}
  	\caption{ \label{cms} Stellar mass and star formation histories of galaxies following the main sequence from $z$$=$5 to $z$$=$0.3. The colored lines corresponds to different mass seeds from 10$^6$ (dark blue) to 10$^{10}$ M$_{\odot}$ (red). The solid lines correspond to range in M$_*$ and SFR constrained by observations while the dashed lines are extrapolations of the MS relation. On the top panel, symbols indicate the time and mass when the galaxies enter the passive zone of the UVJ diagram (see Fig.~\ref{color}).}
\end{figure}

\begin{figure}
  	\includegraphics[width=\columnwidth]{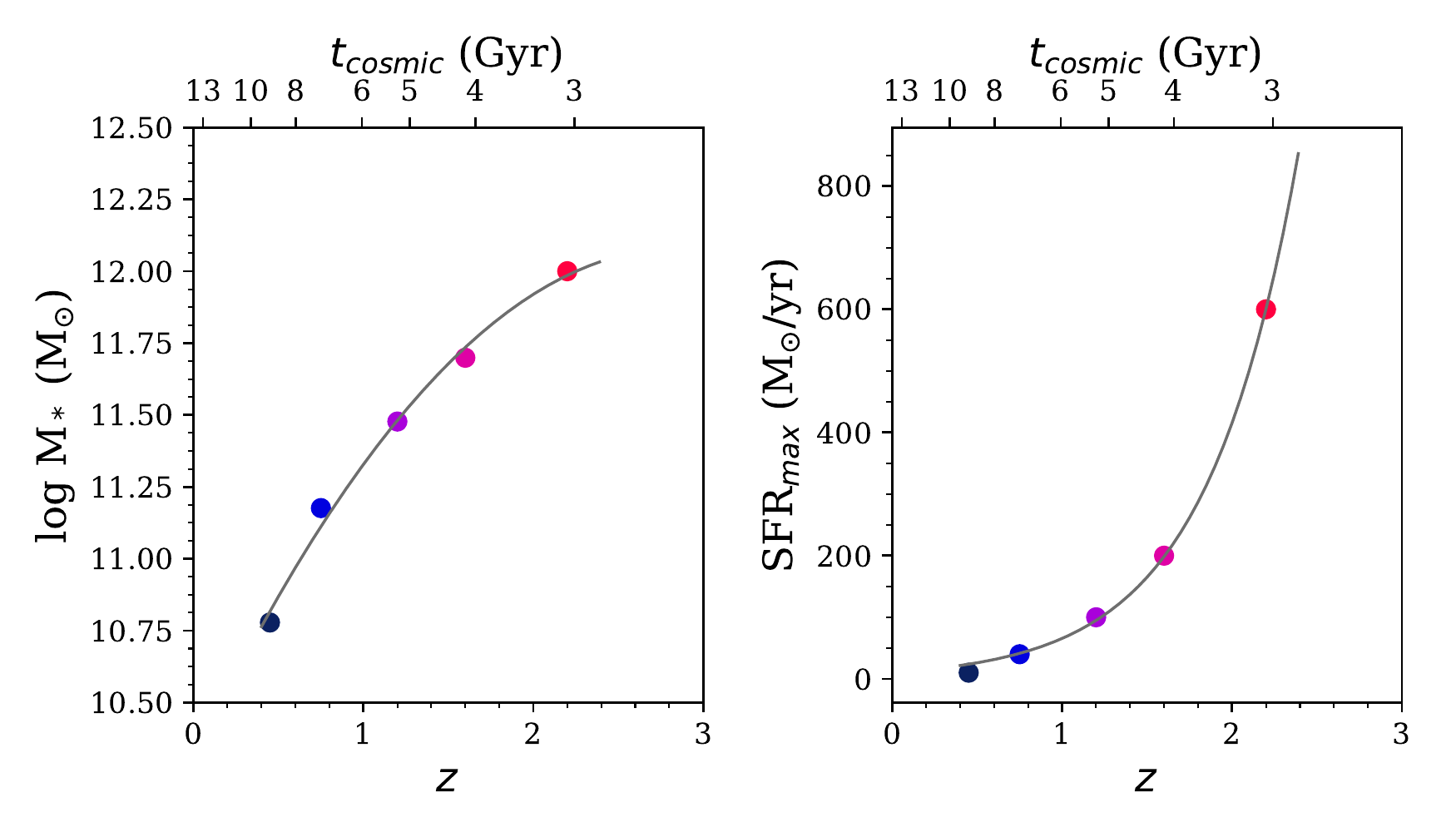}
  	\caption{ \label{halo} Stellar masses (left panel) at which the galaxy growth starts to drastically slow down as a function of redshift and SFR at the peak of the SFH (right panel). Points are color-coded according to their seed mass as defined in Fig.~\ref{cms}. Fit to the data are shown in grey in both panels.}
\end{figure}

We show the resulting SFHs on Fig.~\ref{cms} (bottom panel), from $z$$=$5 to 0.3, for four different seed masses, as well as their stellar mass histories (top panel).
The ranges where the MS relations is constrained from observations, i.e. up to M$_*$$=$3$\times$10$^{11}$, is shown (solid lines) as well as extrapolations of these relations at higher masses (dashed lines).
At early stages, the SFR is increasing with cosmic time implying that the effect of the positive slope of the MS dominates the decrease of its normalization.
Then it reaches a peak, after which, a smooth decline of the SFR is observed due to a combination of the decline of the MS normalization and the bending of the MS at high masses.
The peak of the SFR history of an individual galaxy depends on the mass of the seed and occurs at a cosmic time of 2.8\,Gyr in the case of $M_{seed}$$=$10$^{10}$\,M$_{\odot}$, and at 8.9\,Gyr for $M_{seed}$$=$10$^{6}$\,M$_{\odot}$ (Fig~\ref{halo}, right panel).
From this point, we can note that the position and width of the peak of the cosmic star formation density must contain information on the distribution of galaxies seed masses. 
The SFH of MS galaxies thus depends on the time but also on the seed mass.
This seed mass can be used as a proxy for the halo mass and thus the environment \citep{Behroozi13}.
Therefore, the star formation history of a galaxy following the MS is prone to environment.
These different shapes of the SFH as a function of the seed mass are a direct consequence of the bending of the MS above a given mass, whose evolution with redshift is taken into account by the models of \cite{Schreiber15}.
After the peak, the SFR starts to smoothly decrease translating in a much slower stellar mass growth, close to saturation (Fig.~\ref{cms}, top panel).
Thus, just by following the MS, a galaxy will undergo a smooth and slow diminution of its star formation activity from a time defined by its seed mass.

\begin{figure}
  	\includegraphics[width=\columnwidth]{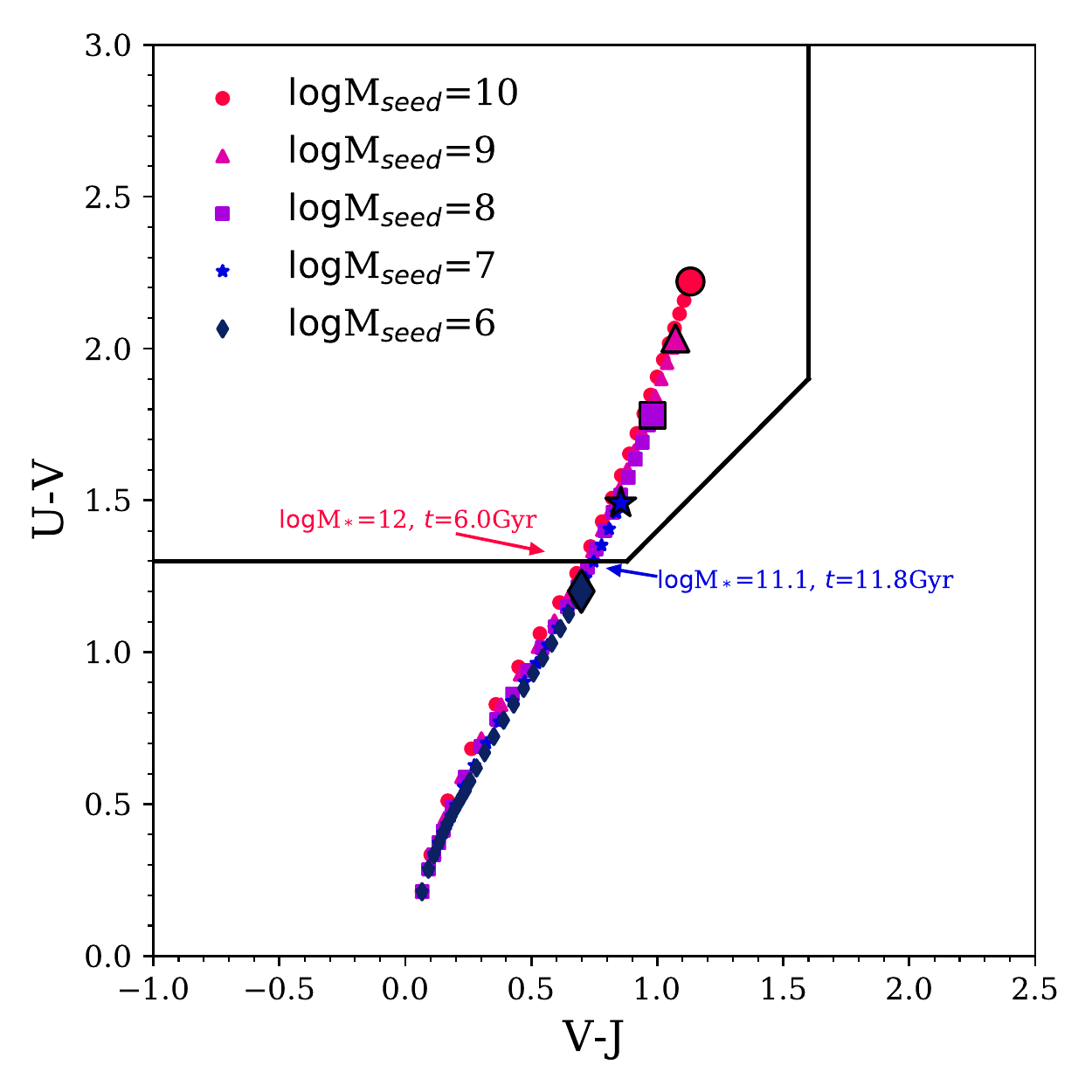}
  	\caption{ \label{color} Evolution of the U-V and V-J colors in the UVJ diagram from $z$=$5$ to $z$=$0.1$ for the five galaxies presented in Fig.~\ref{cms}, assuming the full star formation histories of Fig.~\ref{cms}(constrained + extrapolations) and no dust attenuation. Color coding is the same as in Fig.\ref{cms}. Black contoured symbols indicate the final time step. The time and stellar mass when the galaxies with seed mass of 10$^6$ (dark blue) and 10$^{10}$ M$_{\odot}$ (red) enter the passive ``red'' region are indicated.}
\end{figure}

This is the scenario of the slow downfall presented in \cite{Schreiber15}, for instance.
In Fig.~\ref{color}, we show the evolution of the colors of  the galaxies presented in Fig.~\ref{cms}, assuming no dust attenuation.
Just by following the MS, the evolution of its colors makes the galaxy enter the ``passive'' zone of the UVJ diagram.
However, the galaxy is not quenched, it is still following the MS and thus forming stars, but given the relatively low SFR and the high mass of the galaxy the colors of the evolved stellar population dominate the emission of the galaxy and place it into the ``passive'' region of the UVJ diagram.

Fig~\ref{halo} shows the critical stellar mass and redshift at which the mass growth starts to reach a plateau (left panel) and the SFR at the peak of the SFH (right panel) for the five seed masses shown in Fig.~\ref{cms}.
There is a strong variation of the mass  and SFR$_{max}$ with redshift.
Using abundance matching, one can convert these stellar masses to halo masses \citep{Behroozi13} obtaining 10$^{13}$\,M$_{\odot}$ at $z$$=$0.4 to 2-4$\times$10$^{14}$\,M$_{\odot}$ above $z$$=$1-1.5.
At all redshifts, these corresponding halo masses are above the critical line predicted by \cite{Dekel09} to set the threshold mass for a stable shock based on spherical infall analysis \citep{Dekel06}.
At these mass-redshift regions, the model presented in \cite{Dekel09} predicts that the galaxies lie in hot haloes shutting off most of the gas supply to the inner galaxy.
This resulting limited gas inflow on the galaxy could be a possible explanation for the slow decrease of the star formation activity.

\begin{figure}
  	\includegraphics[width=\columnwidth]{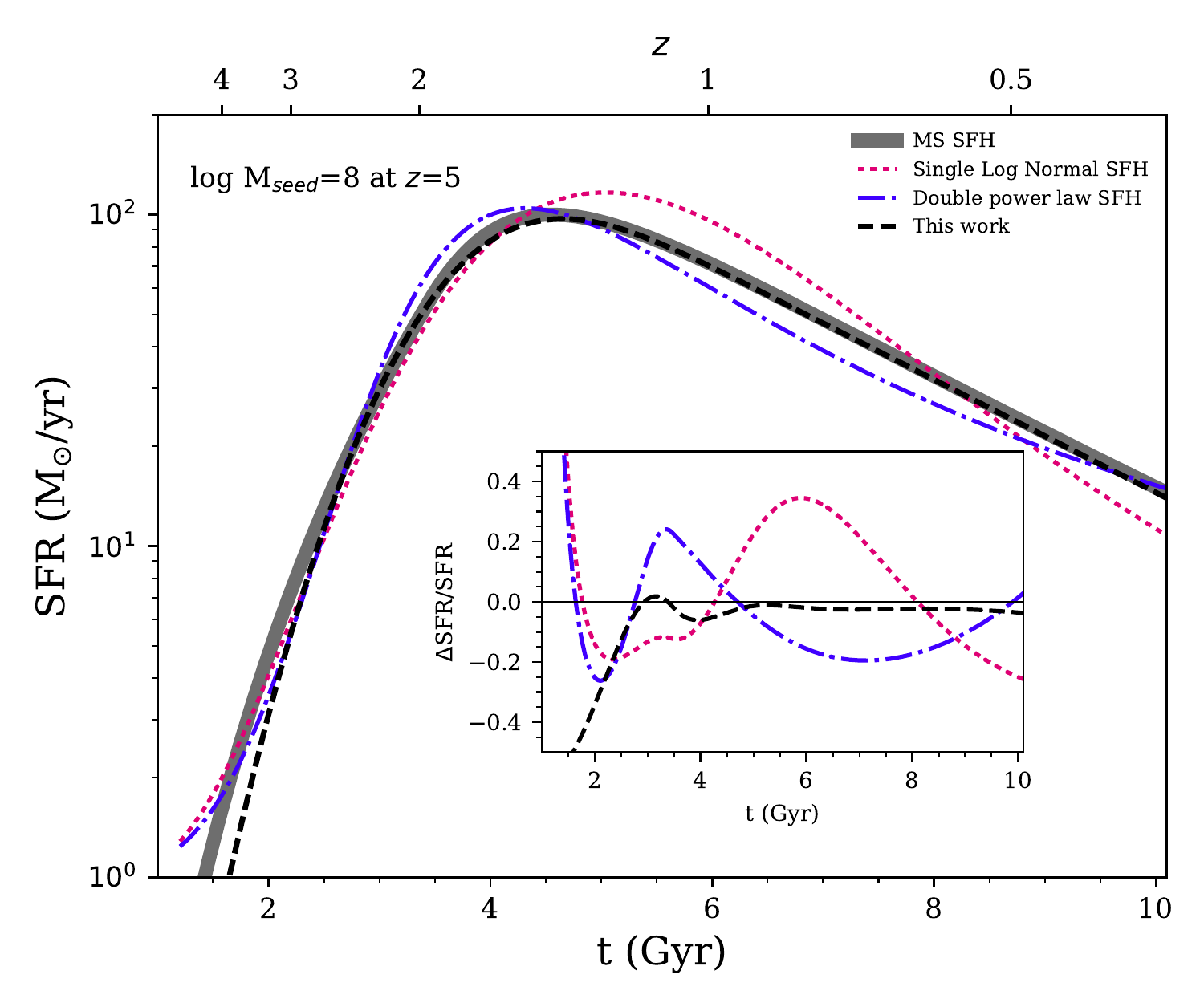}
  	\caption{ \label{func} Fit of the MS SFH assuming a seed mass of 10$^8$\,M$_{\odot}$ at $z$$=$5, using different analytical functions proposed in the literature: log-normal (red), double power law (blue), and the parametrization based on a Right Skew Peak Function proposed in this work (black). The residual of each fit is shown in the insert panel.}
\end{figure}

Several studies aimed at trying to parametrize the global SFH of galaxies with analytical functions, either with the requirement to fit both individual SFH as well as the cosmic SFH, such as one or two log-normal functions \citep{Gladders13,Abramson16}, or to follow the predicted evolution of Dark Matter (DM) halo masses, such as a double power law \citep{Behroozi13}.
These functional forms, however, do not provide satisfactory fits of the computed SFH of MS galaxies, as shown in Fig.~\ref{func}.
These functions do not manage to model the early SFH and the position of the peak of SFR is offset.
In the case of the log-normal SFH, the peak is shifted towards later times and is too broad.
Furthermore, the slope of the declining part is too steep.
The double power law provides a peak slightly shifted towards shorter ages, the slope of the declining part is closer to the true one but the declining part of the MS SFH is flatter than what is computed by these functions.
We also tried other forms such as a Gaussian, a skewed Gaussian, and combinations of them with a power law or an exponential, but they all suffered problems.

After an extensive test of several analytic forms, the best fit of the MS SFH (Fig.~\ref{cms}) is found to be a Right Skew Peak Function, based on a skewed Gaussian, which analytical expression is:
\begin{equation}	
    \mathrm{SFR}(t, M_{seed}) = A\, \frac{\sqrt{\pi}}{2}\, \sigma \, e^{\left((\frac{\sigma}{2\,r_{S}})^2 - \frac{t - \mu}{r_{S}}\right)}\,\mathrm{erfc}\left(\frac{\sigma}{2\,r_{S}} - \frac{t - \mu}{\sigma}\right),
\end{equation}	
\noindent with $t$ the time (in Gyr), $A$ the amplitude, $\sigma$ the width of the Gaussian, $r_S$ the right skew slope, $\mu$ the position of the Gaussian centroid, and $erfc$ is the standard complementary error function.
	
\begin{figure}
  	\includegraphics[width=\columnwidth]{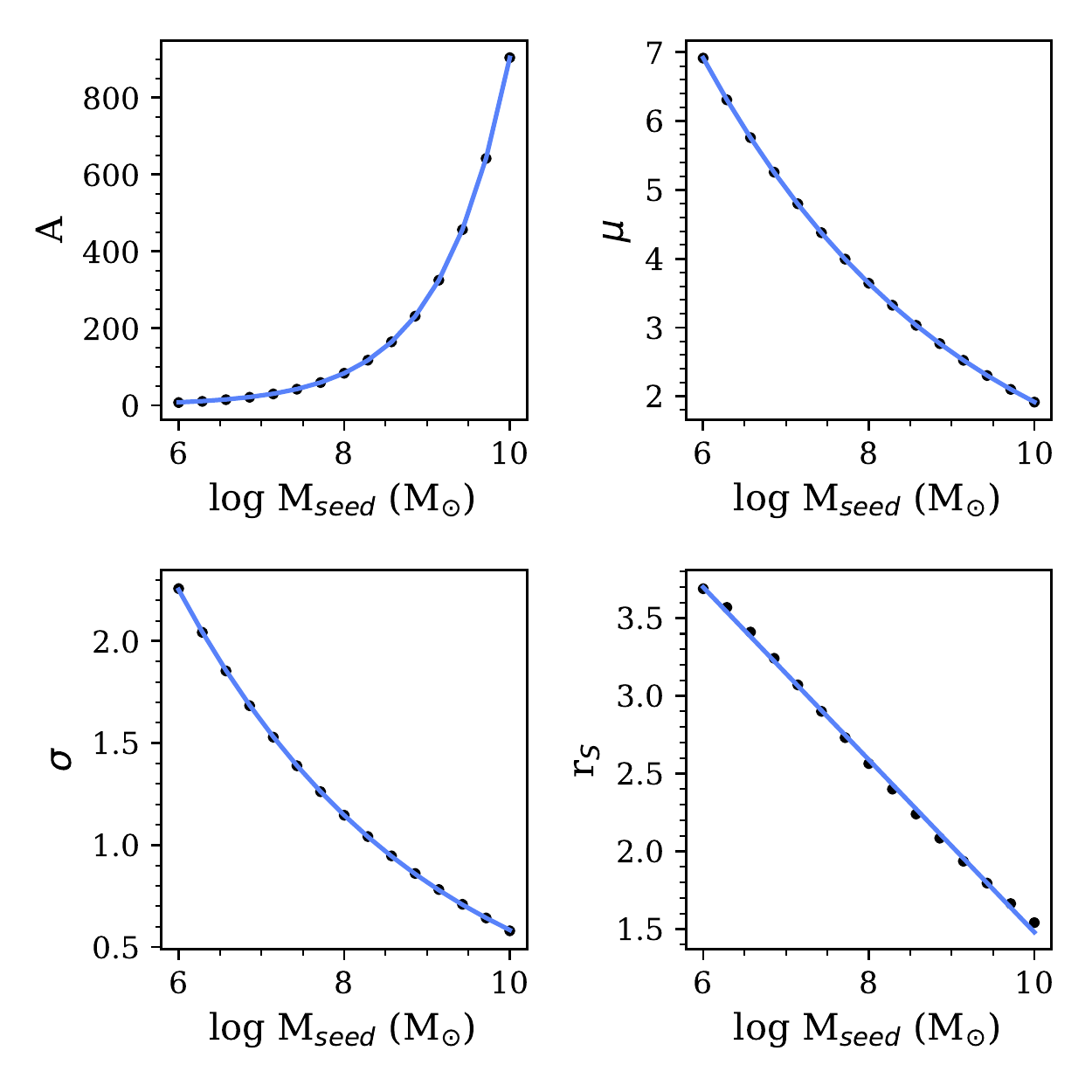}
  	\caption{ \label{paramsfit} Dependency of the Gaussian and exponential parameters obtained from the best fits of a set of SFH, as a function of the seed mass. Black dots are the best values at each seed mass and the red curve shows the relation between the parameters and the seed mass.}
\end{figure}

The values of the different coefficients depend on the seed mass.
To constrain this relation, we fit a Right Skew Peak Function to a set of MS SFH computed from a grid of seed masses.
The best value of each parameter as a function of seed mass is shown in Fig.~\ref{paramsfit} with a best fit for $A$, $\mu$, and $\sigma$ provided by :
\begin{equation}	
	p(M_{seed}) = A_p e^{-\frac{\log M_{seed}}{\tau_p}}.
\end{equation}	
The values of $A_p$ and $\tau_p$ are given in Table~\ref{coef} for all the coefficients. 
For the right skew slope parameter, $r_S$, a linear function provides a better fit with the parameters also provided in Table~\ref{coef}.
The results of this parametrization of the MS SFH of a galaxy with a seed mass of 10$^{8}$\,M$_{\odot}$ is shown on Fig.~\ref{func}.
The early part of the MS SFH is slightly underestimated with the Right Skew Peak Function but at $t$$>$3\,Gyr the agreement is very good, the position of the peak is recovered and the declining slope is reproduced.

\begin{table}
	\centering
	\caption{Coefficient values to compute the parameters of the MS SFH corresponding to an exponential declining function for $A$, $\mu$, and $\sigma$, and to a linear function for $r_S$. }
	\begin{tabular}{ l c c }
	 	\hline\hline
		Parameter &$A_p$ & $\tau_p$ \\ 
		\hline
		$A$ & 6.0$\times 10^{-3}$ &   -0.84 \\
		$\mu$ &    47.39 &    3.12\\
		$\sigma$ &    17.08 &    2.96 \\
	 	\hline\hline
		Parameter &slope: $\alpha_p$ & $\beta_p$ \\ 
		\hline
		$r_S$ &   -0.56 &    7.03\\
		\hline		
		\label{coef}
	\end{tabular}
\end{table}

This SFH is computed from the MS relation obtained from observations.
The bulk of star-forming galaxies follow this relation and thus we use it as a benchmark in the rest of this study.

\section{\label{models}Analytical star formation histories}

\begin{figure*}
	 \includegraphics[width=\textwidth]{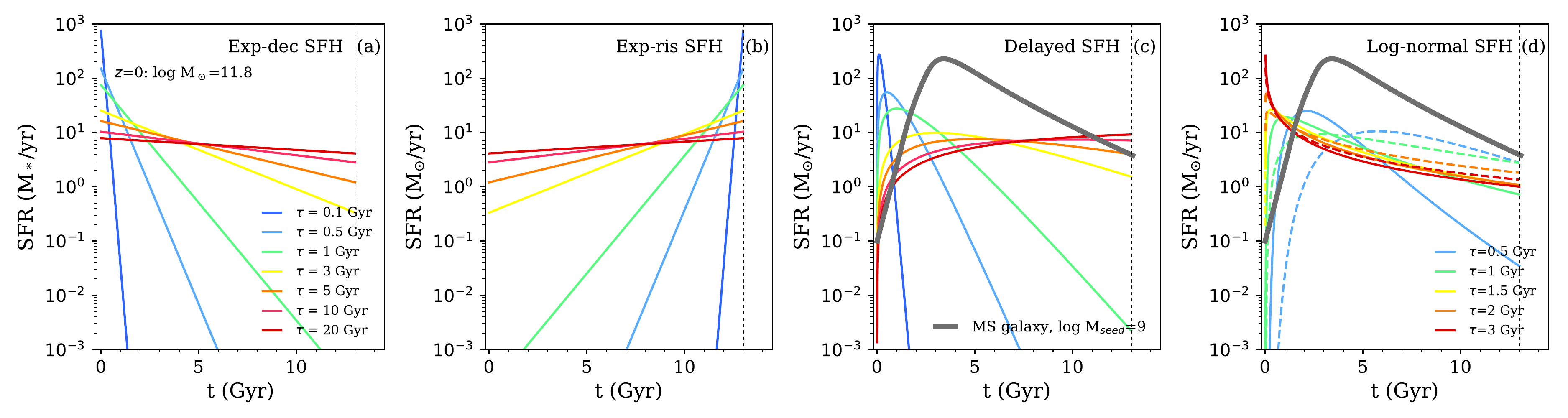}
	  \caption{ \label{sch} From left to right: exponentially declining (a), exponentially rising (b), delayed (c), and log-normal (d) SFH considering a galaxy with a stellar mass of 6.3$\times$10$^{11}$\,M$_{\odot}$ at $z$$=$0. We compare the delayed and lognormal SFH shapes with our MS galaxy SFH in grey (see Fig.~\ref{func}), in the last two panels. For the log-normal SFH, models are shown with $\ln t_0$=1\,Gyr (solid lines) and $\ln t_0$=2\,Gyr (dashed lines).}
\end{figure*}

In this Section, we test the robustness of various analytical forms of SFH, commonly used in the literature.
We focus on the delayed, the exponentially declining, the exponentially rising, and the log-normal SFHs.
We describe their parametrization and show their shape on Fig.~\ref{sch}, as a function of the e-folding time parameter of the main stellar population $\tau$.

The exponential SFH is defined as: 
\begin{equation}
	\mathrm{SFR}(t) \propto e^{-t/\tau}.
\end{equation}
\noindent where $t$ is the time and $\tau$ the e-folding time of the stellar population.
Positive and negative values of $\tau$ corresponds to an exponentially declining and rising SFH, respectively.
	
The delayed SFH is defined as:
\begin{equation}
	\mathrm{SFR}(t)  \propto t \,e^{-t/\tau}.
\end{equation}
\noindent where $t$ is the time and $\tau$ the e-folding time of the stellar population.

The effect of $\tau$ on the shape of the delayed and exponential declining SFHs can be seen on Fig.~\ref{sch}: short values correspond to galaxies where the bulk of the stars was formed early on and in a small time, followed by a smooth decrease of the SFR, while high values imply a roughly constant SFR over cosmic time.
For the exponentially rising SFH, the bulk of the stars is formed in a short and recent time with small $\tau$ values.
For these three SFHs, the highest SFR are reached for low $\tau$ values and short times.
As a comparison, we show along with the different delayed SFH the modeled SFH of a MS galaxy with a seed mass of 10$^{9}$\,M$_{\odot}$ at $z$$=$5.
No value of $\tau$ provides a good model of the peak of the MS SFH, reaching such a SFR would need a low value of $\tau$ associated with a short age incompatible with the time at which the peak of SF occurs.

Recently, \cite{Abramson16} discussed the use of a log-normal function to model galaxies SFH and reproduce the cosmic SFR density following the results of \cite{Gladders13}.
In these studies, the log-normal SFH is defined as:
\begin{equation}
	\mathrm{SFR}(t)  \propto \frac{1}{t \sqrt{ 2 \pi \tau^2 }}\,e^{\frac{-(\ln(t) - t_0)^2} {2  \tau^2}},
\end{equation}
\noindent where $t_0$ is the logarithmic delay time, and $\tau$ sets the rise and decay timescale \citep{Gladders13}.
The log-normal SFH, for different values of $t_0$ and $\tau$, is shown in Fig.~\ref{sch}d.
The inclusion of $t_0$ controls the time at which the SFH peaks, which can not be done in the case of the delayed SFH for instance.
We compare these different log-normal curves with the computed SFH of MS galaxies, as discussed in the previous Section and Fig.~\ref{func}, and find that the log-normal function also struggles to reproduce this SFH (Fig.~\ref{sch}).
While adapted to reproduce the cosmic SFH integrated over all galaxy seed masses, the log-normal SFH is less efficient in reproducing the shape of the individual SFH of MS galaxies.
These histories are indeed strongly dependent on seed masses as illustrated in Fig.~\ref{cms}-bottom.

\section{\label{ms} Recovering the star formation history of main sequence galaxies}

In Sect.~\ref{sfhms}, we built the SFH of a galaxy following the MS during all its lifetime.
To test how well the analytical SFHs presented in the previous Section can model this archetypal galaxy, we show in Fig.~\ref{ms} the specific SFR (sSFR) of this galaxy as a function of its stellar mass, following its evolution on this plane.
In addition, we show the MS relations derived by \cite{Schreiber15} from observations at $z$$=$0, 2, and 4.
This relation is only constrained up to $z$$=$3 but we extrapolate it at $z$$=$4 following \cite{Schreiber17} who recently showed that it still holds at $z$$=4$.
To understand if the analytical SFHs studied here are able to cover the parameter space necessary to model our MS galaxy, we compute a grid of models for each SFH.
We use the largest reasonable limits of age, from 100\,Myr to 13\,Gyr, and $\tau$, from 1\,Myr to 20\,Gyr for the exponential SFHs and the delayed one.
The number density of models is completely dependent on our $\tau$ and age grids, however the space covered by our computed models is the largest since we consider minimum and maximum possible values for these two parameters.
The resulting parameter space in terms of sSFR and stellar mass is shown with the colored filled region on Fig.~\ref{ms}.
In the case of the log-normal form, we use three values of $t_0$, and a less resolved grid of $\tau$ as compared to the other models for computational reasons, although we keep the minimum and maximum possible values.

We have seen that the evidence of a MS implies that galaxies experienced on average a rising SFH since their birth before reaching a peak and a drop of star formation.
As a natural result, the exponentially declining SFH is not suited for these early phase.
The locus of the $z$$=$4 MS indeed falls in the dark zone of Fig.~\ref{ms}a, which is unpopulated by the analytical formula whatever parameters are chosen.
More surprisingly, the delayed SFH, which was built to account for this behavior encounter a similar problem although less sharply, i.e., some extreme parameter choices can marginally reach $z$$=$4 but not above.
This comes from the stiffness of the analytical formula, unable to follow the actual shape of the MS driven SFH.
More generally, these two analytical SFH are not optimized to finely sample the SFH at epochs $z$$\geq$2.
This problem is not present with the exponentially rising SFH which instead is not optimized for $z$$\leq$2.
The same conclusions are drawn for the log-normal function that is able to reach very high values of SFR, but not the smallest ones expected at low redshifts for massive galaxies.

\begin{figure*}
  	\includegraphics[width=\textwidth]{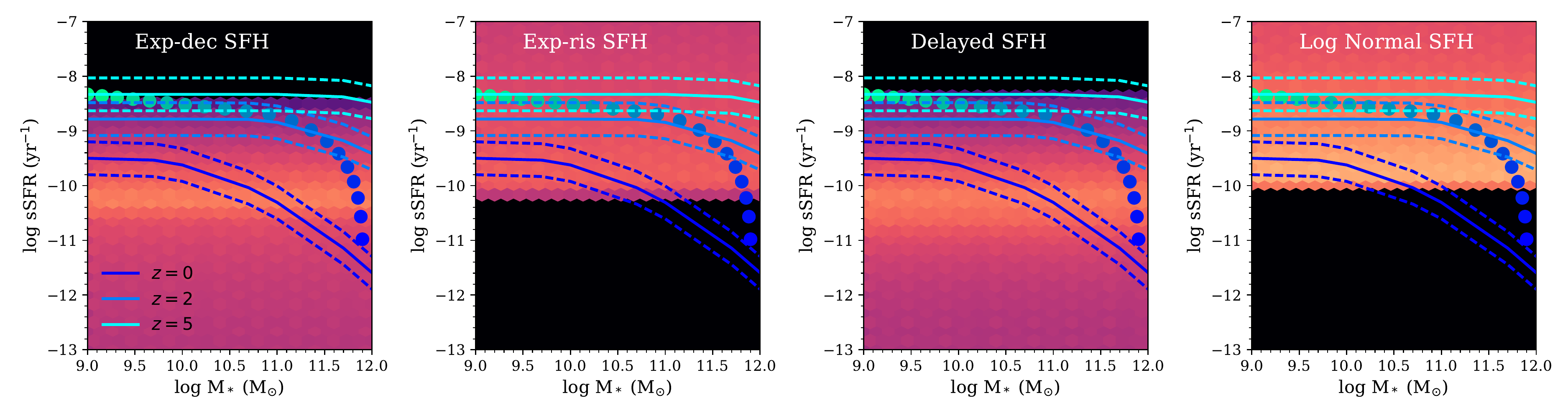}
  	\caption{ \label{ms} Parameter space covered by the exponentially decreasing, exponentially rising, and delayed SFH in the sSFR-M$_*$ plane. The minimum age considered is 100\,Myr and the minimum $\tau$ considered is 1\,Myr. The color shaded regions indicates the number density of models, black representing the absence of models whereas orange indicates the density of models. In each panel, the solid line and the dashed lines show the MS relation and its scatter as derived by \cite{Schreiber15} at $z$$=0$ (dark blue), 2 (light blue) and 4 (cyan). The evolution of a MS galaxy throughout its cosmic life from $z$$=$5 to 0 is shown by the circles color-coded with the redshift.}
\end{figure*}

\begin{figure*}
  	\includegraphics[width=\textwidth]{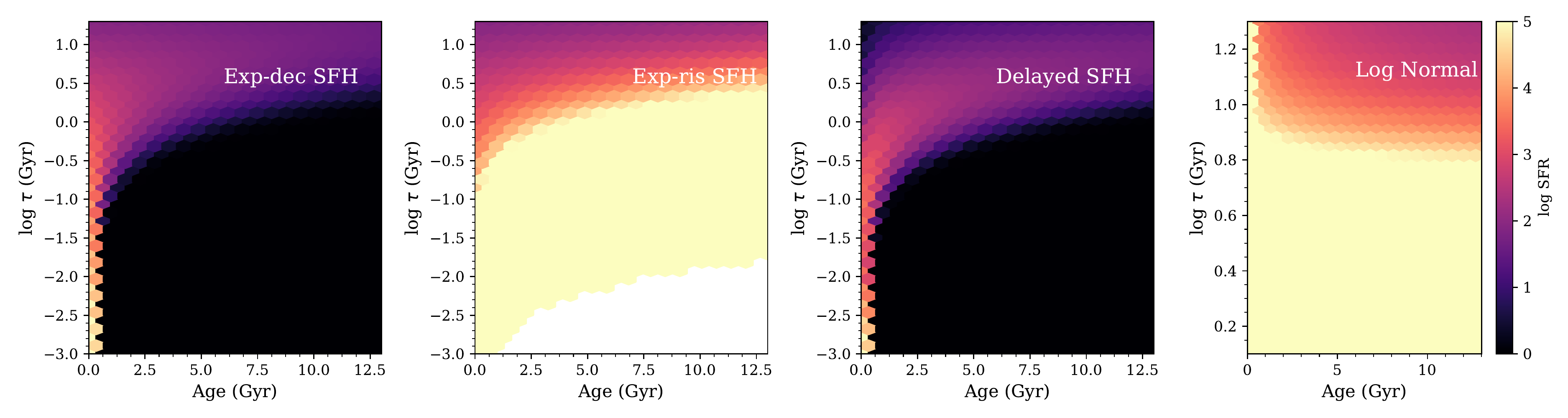}
  	\caption{ \label{tauvage} The dependency between the age and $\tau$ parameters color-coded with the SFR obtained for our grid of star formation histories at $z$$=0$, 2, and 5. The highest SFR are obtained with a combination of low values of $\tau$ but especially the lowest values of the age.}
\end{figure*}

To understand why the exponentially declining and delayed SFHs struggle to reproduce high sSFR, we show in Fig.~\ref{tauvage} the relation between the age, $\tau$, and the SFR for the whole grid of models.
The highest SFRs are reached from a combination of low ages and low values of $\tau$.
This implies limitations on the estimate of the SFR since the age can not be higher than the age of the Universe.
On the contrary, we can easily see that a whole range of high SFRs can be obtained with the exponentially rising SFH and the log-normal one but not the low SFRs needed to model high mass local galaxies, confirming what we observe in Fig.\ref{ms}, where this SFH fails to reproduce our modeled MS galaxy at $z<0.4$.

We conclude that the exponentially declining and delayed SFH are not optimized to model the emission of MS galaxies at $z$$\geq$2 as they show little mathematical flexibility to reproduce the SFR of these galaxies.
However the exponentially rising SFH can model such high redshift sources, as well as the log-normal function, but not for massive galaxies below $z$$<$2.

\section{\label{scatter}Recovering the scatter of the main sequence}

The MS relation exhibits a scatter of 0.3\,dex that is found to be constant with redshift \citep{Guo13,Schreiber15}.
There are two possible origins for this scatter.
The first one is that this scatter could be artificially created by the accumulation of errors due to observations, photometric measurements, and determination of physical properties from different methods (instrumental errors, assumptions linked to SED modeling, etc).
The second one is that the scatter could be due to a variation of a third property (other than M$_*$ and SFR) across the MS and thus be physical \citep[e.g.,][]{RodriguezPuebla16}.
Indeed, some simulations predict that galaxies can undergo some fluctuations of star formation activity resulting in variations of their SFR such as compaction or variations of accretion \citep[e.g.,][]{DekelBurkert14,Sargent14,Scoville16}. 
These variations should be small enough to keep the SFR of the galaxy within the MS scatter

To understand the origin of the scatter of the MS, analytical SFHs must be able to recover small recent variations of the SFR with a precision better than the scatter of the MS itself, i.e. 0.3 dex.
To test the functional forms studied in this work, we take our computed MS galaxy SFH described in Sect.~\ref{ms} and add, at different redshifts, a small variation of the SFR, an enhancement or a decrease.
This variation is applied to the last 100\,Myr of the SFH and its intensity is randomly selected in a Gaussian distribution centered on the exact MS SFR at the given redshift and mass, and with a $\sigma$ of 0.3\,dex.
Five hundreds mock SFHs are created at each redshift, from $z$$=$1 to 5, varying the mass of the galaxy seeds from 10$^7$ to 10$^{10}$ M$_{\odot}$.
To derive galaxies SED from these SFH, we use the SED modeling code CIGALE \citep[][Boquien et al. in prep]{Noll09} based on a energy balance between the energy attenuated in UV-optical and re-emitted in IR by the dust.
CIGALE can model galaxies SED from all kind of SFH, analytical forms or output from complex simulations.
In this case, we provide CIGALE with our computed SFHs and build SEDs using the stellar population models of \cite{BruzualCharlot03}, a \cite{Calzetti00} attenuation law, and the dust emission templates of \cite{Dale14}.
The modeled SEDs are then integrated into the set of filters available for the GOODS-South field, and a random noise distributed in a Gaussian with $\sigma=$0.1 is added to the modeled fluxes, following the method described in \cite{Ciesla15}.
We then use the SED fitting mode of CIGALE to recover the physical parameters of our mock galaxies.
To fit our galaxies, we do not provide as an input the parameter used to build the mock galaxies, and the attenuation law is a free parameter.
However, we use three infrared (IR) bands, MIPS 24\microns, PACS 100 and 160\microns.
Because CIGALE is based on an energy budget, having IR data is thus ideal to constrain the amount of attenuation and break the degeneracy with age.
We emphasize the fact that the following results should be considered as a best case scenario. 
The physical properties of the galaxies are computed from the probability distribution function (PDF) of each parameter computed by CIGALE, the final value being the mean of the PDF and the error its standard deviation.

\begin{figure}
  	\includegraphics[width=\columnwidth]{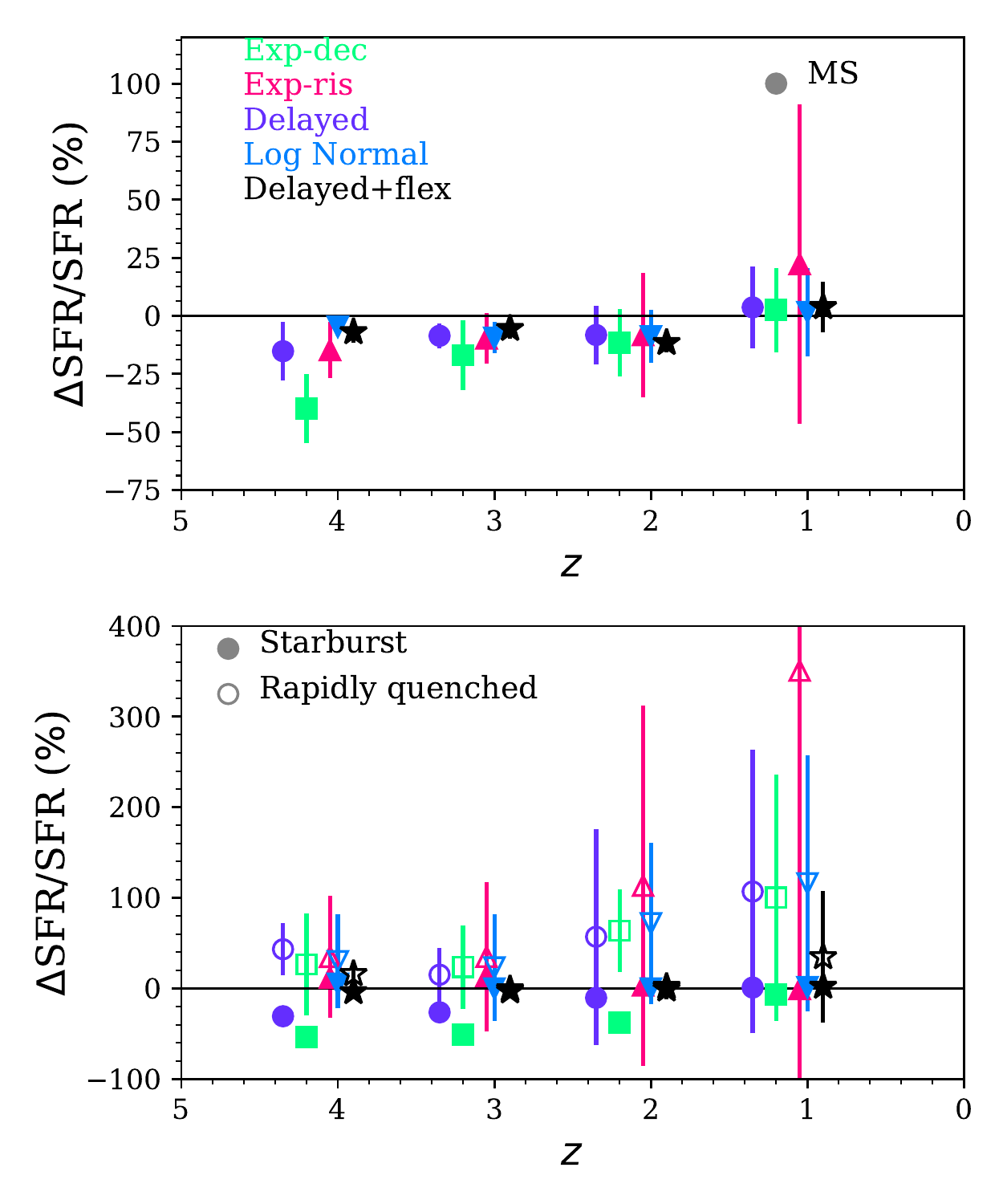}
  	\caption{ \label{mscattermean} Mean errors ($\Delta\mathrm{SFR}/\mathrm{SFR}$=$(\mathrm{SFR}-\mathrm{SFR_{true}})/\mathrm{SFR_{true}}$) on the recovering of the SFR of simulated MS galaxies (top panel), starbursts, and rapidly quenched galaxies (botton panel) as a function of redshift using different SFH assumptions: exponentially-declining (green), exponentially-rising (pink), the delayed (purple), the log-normal (blue), and the delayed + flexibility SFH (black) that is described in Section~\ref{solution}, and at four redshifts. The points are slightly shifted horizontally with respect to each other for the sake of clarity.}
\end{figure}

At each redshift, and for each SFH assumption, Fig.~\ref{mscattermean} (top panel) shows the relative difference between the estimated SFR and the true ones known from our simulated galaxies.
Globally, at all redshifts, all SFH assumptions recover the true SFR within an error below $\pm$25$\%$ except the exponentially declining SFH that underestimates the SFR by $\sim$40$\%$ at $z$$=$4.
The estimated error of 25$\%$ on the SFR corresponds to a quarter of the MS scatter, thus the usual SFH should be sensitive enough to probe variations of the order of 0.3\,dex of star formation activity around the MS.

The ability of each SFH assumption to recover the physical properties of the mock galaxies is however dependent on redshift.
Indeed, the error on the SFR  is increasing with redshift for the exponentially declining SFH showing that this form is more suited for galaxies at $z$$\leq$2-3.
The opposite behavior is observed for the rising exponential providing better estimates of SFR at $z$$=$4 than at $z$$=$1 where it clearly overestimates this parameter, indicating that this SFH is more suited for galaxies at $z$$>$2-3.
This is in agreement with the typical MS SFH described in Sect.~\ref{ms} with an increase of the SFR at early times, well modeled by the rising exponential, followed by a smooth decrease, modeled by the exponentially declining SFH.
The behavior of these two SFH assumptions is not surprising and was already pointing out in several studies \citep[e.g.,][]{Maraston10,Papovich11,Pforr12,Reddy12}.
However, this emphasizes the fact that these two SFH must be carefully selected according to the redshifts of galaxies in order to limit biases on the derivation of their physical properties.
The delayed SFH however seems to provide estimates of the SFR showing a weaker dependency on redshift although we note an underestimate at $z$$=$4 that is less pronounced than the exponential SFH.

Recently, \cite{Ciesla15} performed a similar analysis on $z$$=$1 galaxies simulated by the semi-analytical code GALFORM and found that the exponentially declining and delayed SFH were underestimating the SFR by 11$\%$ and 9$\%$, respectively.
Here we find a weak overestimation of a few percent in the estimate of the SFR for both SFH.
This apparent disagreement is likely due to different SFH used to built the SEDs (complex SFH produced by a SAM code in \cite{Ciesla15} versus a simple analytical form linked to the MS here) as well as the parameters used to perform the fits, for instance the attenuation law is a free parameter in this study.
Improvements in the code since 2015 may also play a role.
The log-normal SFH provides very good estimates of the SFR of MS galaxies at all redshifts with an error less than 10\%.
We note that the two functional forms that have similar global shape compared to the MS SFH, the log-normal and delayed SFH, recover well the SFR of the mock galaxies with little dependency on the redshift compared to the exponential forms.
From this analysis, we conclude that the four tested SFH assumptions provide fair measurements of the SFR of MS galaxies, but the log-normal and delayed SFH estimates are more accurate and less redshift dependent than the other models.

\section{\label{outliers}Extreme SFHs: Starbursts and rapidly quenched galaxies}

We tested in the previous Section the ability of the usual SFHs to recover the SFR of MS galaxies. 
Recently, \cite{Steinhardt17} argued that considering that galaxies stay on the MS all their lives would result in a significantly steeper stellar mass function towards low redshift and showed that taking into account mergers would settle the conflict with the observed growth of stellar mass.
Furthermore, after spending a long time on the MS, galaxies are expected to quench \citep[e.g.,][]{Heinis14,SteinhardtSpeagle14}.
To test the impact of higher perturbations on the SFR, we perform here the same test than in the previous Section, i.e. we model MS galaxies to which we apply this time a strong burst or quenching.

To model starburst galaxies, we systematically multiply by a factor of 5.24 the SFR in the last 100\,Myr of our simulated MS galaxies and add a random scatter of 0.3\,dex following a Gaussian distributions, as suggested by \cite{Schreiber16}.
We then follow the same procedure than for the MS galaxies, i.e. we build their SEDs with CIGALE and performing the fitting.
The relative difference between the output of the fitting and the true SFRs for the starburst galaxies are shown in Fig.~\ref{mscattermean} (bottom panel, filled symbols).
At $z$$=$1, the three assumptions provide good estimates of the SFR of starbursts galaxies.
Moreover, the results from the rising exponential SFH are better than for the MS galaxies as expected from Fig.~\ref{tauvage} where we showed that this SFH is able to reproduce very high SFR, and provide good estimates, below 20$\%$, at all the redshifts considered here.
The log-normal function provides very good estimates as well, as expected.
However, the delayed and exponentially declining SFHs underestimate the SFR with a factor increasing with redshift, up to errors of $\sim$40 and $\sim$50$\%$, respectively, at $z$$=$4.
This is also explained by the difficulty that these models have in reaching very high SFR as shown again in Fig.~\ref{tauvage}.
We thus conclude that exponentially rising and log-normal SFHs are well suited to model starburst galaxies at all redshifts, whereas the delayed SFH is suited for starburst galaxies at $z$$\leq$2 and the exponentially declining for galaxies at $z$$\leq$1.

We now investigate the case of galaxies undergoing a fast quenching of their star formation activity.
Here, we want to probe processes that strongly affects the activity of the galaxies in less than 500\,Myr, such as ram pressure striping for instance, or violent negative feedback (e.g. SF-driven or AGN-driven winds) and not smooth quenching occurring in timescales of the order of several Gyr.
Following the results of \cite{Ciesla16}, we apply to our MS SFH a systematic instantaneous break on the last 100\,Myr of the SFH, making the SFR dropping to 15$\%$ of the SFR before quenching.
We then add a random scatter of 0.3\,dex as we did for MS and starburst galaxies.
The value of 15$\%$ is chosen as being intermediate between a total quenching (0$\%$) and 30$\%$ which is the upper limit empirically defined by \cite{Ciesla16} to consider a galaxy as rapidly quenched.
The results on the SFR recovering for the rapidly quenched galaxies are shown in Fig.~\ref{mscattermean} (bottom panel, open symbols).
In this case, the dispersion on the measurement is the largest, all the SFH overestimate the SFR in the best case (delayed SFH at $z$$=$3), by 20$\%$, and by 350$\%$ in the worst case (the exponentially rising SFH at $z$$=$1).
The dependance with redshift of $\Delta SFR/SFR$ is the same for the three assumptions, i.e. they overestimate by 20 to 40$\%$ the SFR of galaxies at $z$$=$3 and 4, but by a factor of 2 in average at $z$$=$1 and 2.
Indeed, we see in Fig.~\ref{tauvage} that the delayed and the exponentially declining SFH can reach very low values of SFR but need a long time/age to reach these values, which is not compatible with a sharp decline of the SFR.
For the exponentially rising SFH, low values of SFR ($<$10\,M$_\odot$yr$^{-1}$) are not reachable as seen in Fig.~\ref{tauvage}, as well as for the log-normal SFH in the times probed here.
We conclude then that none of these four SFHs is suited to derive the SFR of galaxies undergoing a rapid quenching of their star formation activity.

\section{\label{star} Recovering the stellar mass of galaxies}
\begin{figure}
  	\includegraphics[width=\columnwidth]{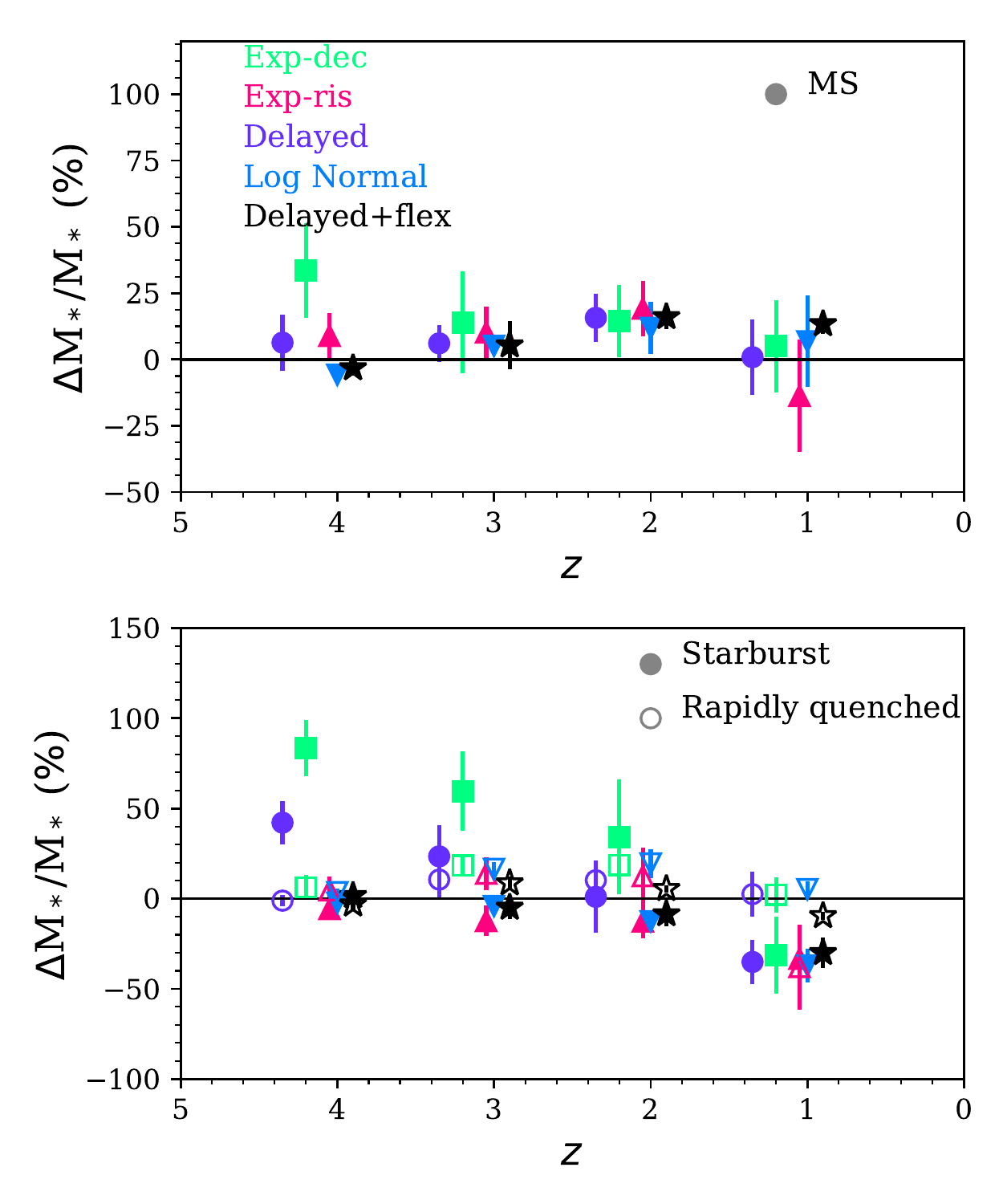}
  	\caption{ \label{mscattermean_mstar} Mean errors ($\Delta\mathrm{M_*}/\mathrm{M_*}$=$(\mathrm{M_*}-\mathrm{M_*^{true}})/\mathrm{M_*^{true}}$) on the recovering of the M$_*$ of simulated MS galaxies (top panel), starbursts, and rapidly quenched galaxies (botton panel) as a function of redshift using different SFH assumptions: exponentially-declining (green), exponentially-rising (pink), the delayed (purple), the log-normal (blue), and the delayed + flexibility SFH (black) that is described in Section~\ref{solution}, and at four redshifts. The points are slightly shifted horizontally with respect to each other for the sake of clarity.}
\end{figure}

The previous discussions on the ability of analytical SFH assumptions to recover the SFR of galaxies were motivated by the fact that at a given stellar mass the range of SFR reachable for these SFHs can be limited for mathematical reasons.
However, we test in this section their ability to recover the stellar mass of the galaxies as well.
We compare in Fig.~\ref{mscattermean_mstar} the M$_*$ obtained from our SED fitting procedure to the true M$_*$ of the galaxies, as we did for the SFR in Fig.~\ref{mscattermean}. 
For MS galaxies, Fig.~\ref{mscattermean_mstar} (upper panel) shows that in almost all cases, i.e. all SFH assumptions and all redshift, the stellar mass is recovered with an error lower than 25\%.
The only exception is for the exponentially declining SFH at $z$$=$4 where M$_*$ is overestimated by 35\%.
For the starburst galaxies (bottom panel of Fig.~\ref{mscattermean_mstar}), there is a relation between the mean error on the stellar mass and the redshift for all SFH assumptions.
The worst case is the exponentially declining SFH with a strong trend with redshift, from an overestimation of nearly 100\% at $z$$=$4 to an underestimation of $\sim$30\% at $z$$=$1.
The same tendency is observed in the case of the delayed SFH with a lower gradient, from $+$40\% to $-$30\%.
The lognormal and exponentially rising SFH provide very good estimate of the stellar mass of starburst galaxies down to $z$$=$1 where the reach an underestimation of 30\% like the exponentially declining and delayed SFH.
In the case of rapidly quenched galaxies, all the SFH assumptions tested in this work provide a very good measurement of their stellar mass, implying that the larger errors observed for the starburst galaxies rise from the need to reach high SFR at a given stellar mass.
We conclude that only the exponentially declining SFH might introduce errors in the measurement of M$_*$, especially in the case of starburst galaxies, and must thus be used cautiously.

\section{\label{bias}Biases}
\begin{figure*}
  	\includegraphics[width=\textwidth]{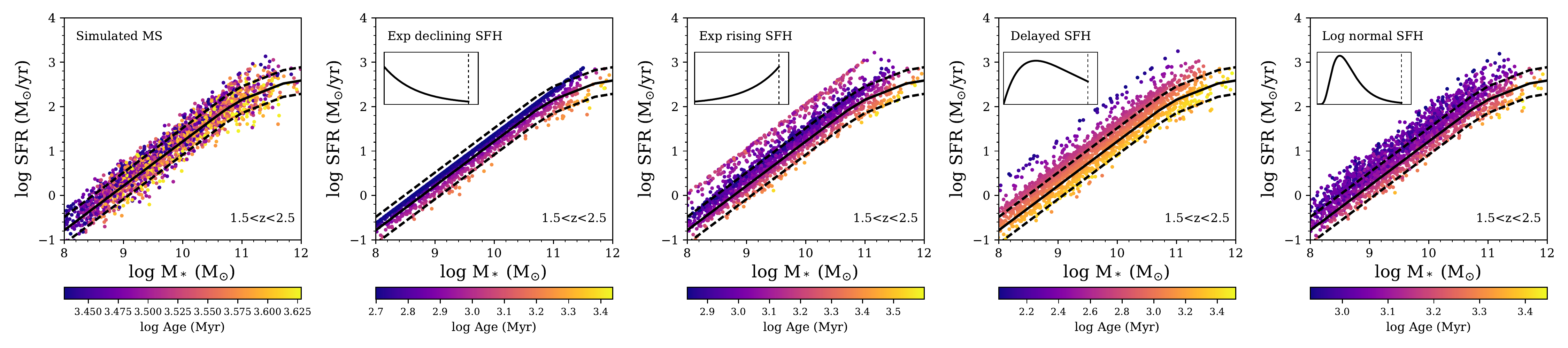}
  	\caption{ \label{mssimu} Left panel: Mock galaxies between $z$$=1.5$ and 2.5 placed on the MS diagram and color-coded according to their age. The solid line and the dashed lines indicate the MS and its scatter at $z$$=2$. From the second left panel to the right panel: Results of the SED fitting of the mock galaxies using, from left to right, an exponentially declining, an exponentially rising, a delayed, and a log-normal SFH. Points are color-coded with the output age produced by the SED fitting. The $z$$=2$ MS from \cite{Schreiber15} is indicated in each panel.}
\end{figure*}

\begin{figure}
  	\includegraphics[width=\columnwidth]{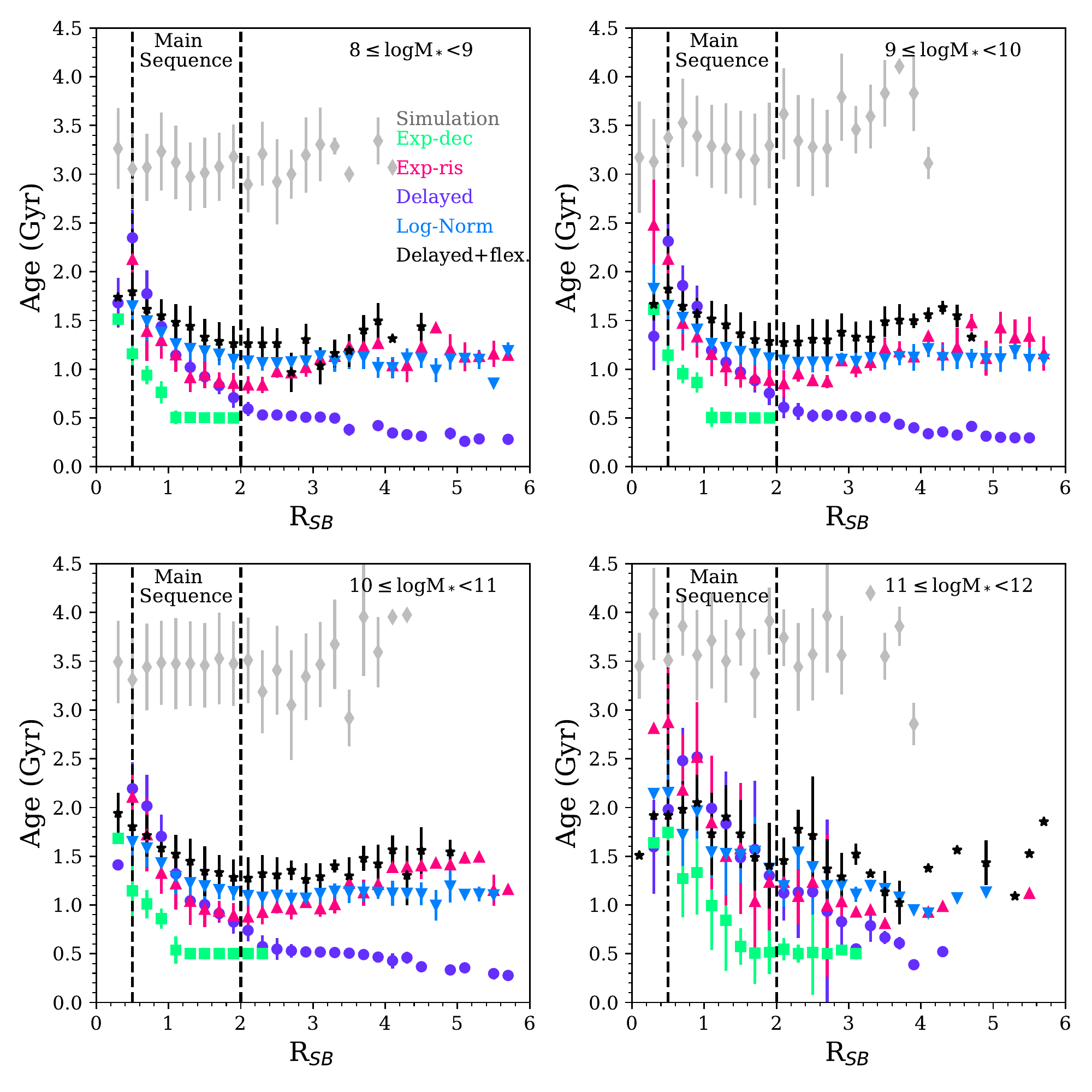}
  	\caption{ \label{mssimursb} Median age of the mock galaxy sample as a function of their distance to the MS (R$_{\mathrm{SB}}$=$\mathrm{SFR}/\mathrm{SFR_{MS}}$) for four different mass bins. The true values of the mock galaxies are shown in grey while the results of the SED fitting using the exponentially declining, the exponentially rising, the delayed, the log normal, and the delayed plus flexibility SFH are indicated in green, pink, purple, light blue, and black, respectively. The error bars indicate the standard deviation around the medians. The dashed lines indicate the limits of the MS.}
\end{figure}

\begin{figure*}
  	\includegraphics[width=\textwidth]{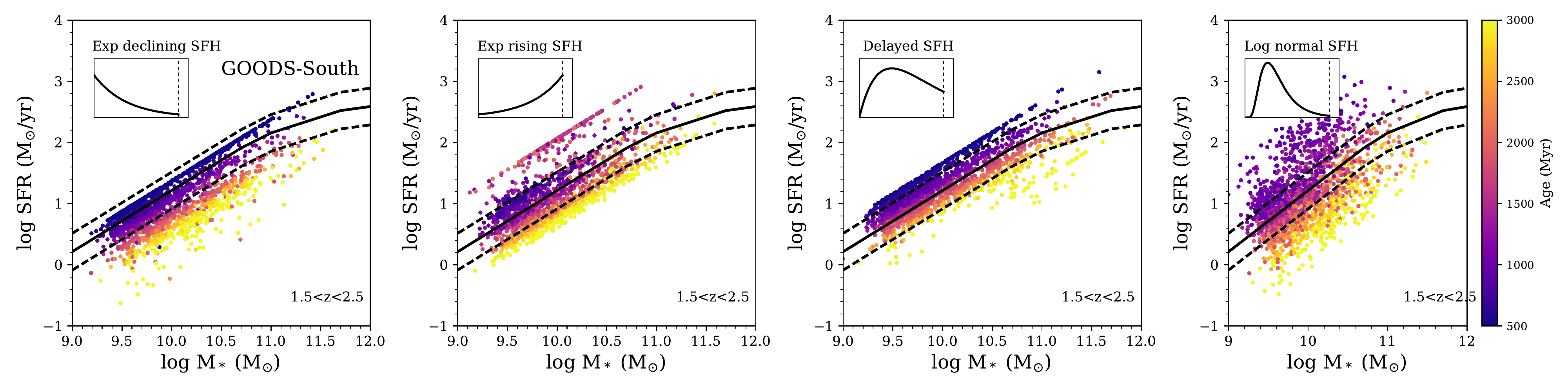}
  	\caption{ \label{msz2} Main sequence relation of GOODS-South galaxies between redshift 1.5 and 2.5 obtained using, from left to right, the delayed SFH, an exponentially decreasing SFH, an exponentially increasing SFH, and a log-normal SFH, color-coded by the age obtained from the fitting procedure. The black line indicates the position of the MS obtained by \cite{Schreiber15} at $z$$=2$.  }
\end{figure*}

In the previous sections, we showed that analytical functions typically used in the literature feature mathematical rigidity that lead to errors in the recovery of physical parameters such as the SFR.
On the one hand, from Fig.~\ref{tauvage}, we understand that large values of SFR are only reached assuming a very short age of the oldest star of the galaxies for the delayed and exponentially declining SFH.
On the other hand, we expect the exponentially rising SFH to struggle in recovering weak SFR.
We thus expect that modeling high redshift galaxies SED with these SFH results into biases, especially in terms of age.
As a first test, we simulate a mock galaxy sample and follow their SFH up to $z$$=$1.5-2.5.
Here we take a simple scenario where a galaxy, with a given M$_{seed}$ at $z$$=5$ stays on the MS all its life but undergoes some episodes of enhanced star formation followed by a time of lower star forming activity, corresponding to a self-regulation.
However these episodes are not strong enough to place the galaxy in the SB zone or in the quiescent region of the MS diagram, we force them to stay within the scatter of the MS.
To do so, we take several assumptions.
First, we randomly pick a M$_{seed}$ in the mass function distribution computed by \cite{Schreiber16} at $z$$=$3.5-4.5.
The normalization of the mass function is not important, only the shape is useful as we normalize it in order to compute the probability of having a galaxy seed with a given stellar mass. 
Then, for each time step we set the probability to have a star formation enhancement to 75$\%$ and, in the case of this small burst, we impose that a small decrease of star formation must follow this enhancement to account for self-regulation.
The intensity of these episodes are randomly chosen in a Gaussian distribution with $\sigma=0.3$\,dex, i.e. within the scatter of the MS.
The intensities of the enhancement and the quiet periods are allowed to be different.
We then calculate at each time step the stellar mass associated to the SFR.
Finally, we randomly pick a redshift between 1.5 and 2.5 as the observed redshift for each galaxies.
In the SED fitting methods, the age is defined as the time where the first star is formed.
In this simple model, using this definition is difficult, we thus arbitrary define the age of the galaxy as the time since the galaxy reached 10$^6$\,M$_{\odot}$.
This assumption should not impact our result since we are interested in relative differences in the age of galaxies and not on absolute value.
The resulting MS is shown in the left panel of Fig.~\ref{mssimu}a.
No SED fitting is performed, this is just the output of the SFH of the mock galaxies. 
These simulated sources span a range of stellar masses between 10$^8$ and 10$^{12}$\,M$_{\odot}$ and SFR between 0.1 and 500\,M$_{\odot}/$yr$^{-1}$.
Their age is comprised between 1.5 and 4\,Gyr.
Younger galaxies are preferentially found at low masses whereas massive galaxies at the highest mass range.
We emphasize here that this is a simple case were the same attenuation law \citep{Calzetti00} and attenuation amount (E$_{(B-V)}$=0.3) is applied to all galaxies.
The only parameters that are different from one simulated galaxy to another are those linked to the SFH.

We apply our SED fitting method to the simulated galaxies and place them on the MS plane using the outputs of the fit for the four analytical forms, color-coded with the age, also derived by CIGALE (Fig.~\ref{mssimu} from panel b to e).
For each SFH, we clearly see that the fitting method artificially introduces an age gradient parallel to the MS, with young ages on the top and older age in the bottom.
The exponentially declining SFH tends to tighten the galaxies on the MS relation while the opposite is noticed for the log-normal SFH that shows a wider spread of the galaxies, especially towards high SFR.
In addition, we note an artificial line at high SFR in the case of the exponentially rising SFH.
While the exponentially declining and rising SFH and the log-normal SFH tend to prefer low values of age, the delayed SFH is producing older ages, compared to the simulated sample.

In Fig.~\ref{mssimursb}, we show the true median age of the simulated galaxies as well as those obtained through SED fitting for all analytical forms studied in this work, as a function of the relative distance to the MS, i.e. the starburstiness \citep[R$_{\mathrm{SB}}$=$\mathrm{SFR}/\mathrm{SFR_{MS}}$,][]{Elbaz11}.
The true age is never recovered by any of the analytical forms studied in this work with underestimations by a factor of 2-3.
Furthermore, while the true age is showing no relation with R$_{SB}$, the age produced by SED fitting shows a clear relation with R$_{\mathrm{SB}}$ with a slope and spread (as indicated by the error bars)  that depends on the SFH assumptions.
The delayed, the exponentially decreasing, and exponentially rising SFHs show the steepest slope, thus are the most biased in all stellar mass bins while the log-normal SFH shows the weakest trend.
We note a tendency to overestimate the R$_{\mathrm{SB}}$ in all the SFH assumptions but the exponentially declining, confirming what we observe in Fig.~\ref{mssimu}.

\begin{figure*}
  	\includegraphics[width=\textwidth]{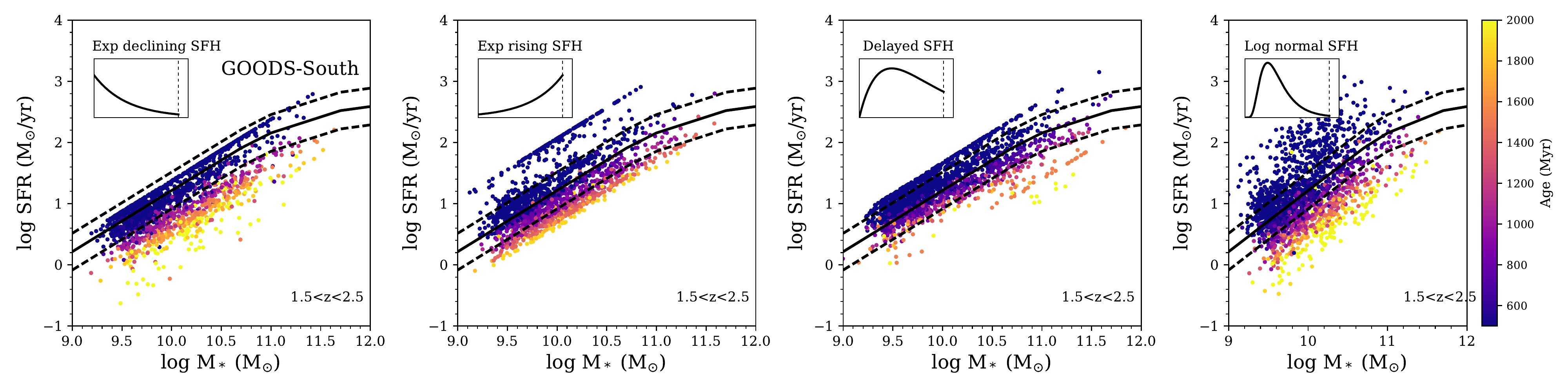}
  	\caption{ \label{msz2mwa} Main sequence relation of GOODS-South galaxies between redshift 1.5 and 2.5 obtained using, from left to right, the delayed SFH, an exponentially decreasing SFH, an exponentially increasing SFH, and a log-normal SFH, color-coded by the mass-weighted age obtained from the fitting procedure. The black line indicates the position of the MS obtained by \cite{Schreiber15} at $z$$=2$.  }
\end{figure*}

The trends observed in Fig.~\ref{mssimu} are obtained in an ideal case where, as we explained, only the SFH varies from one galaxy to another.
To understand how it could impact real data, we select star forming galaxies in the GOODS-South sample from their position in the U-V vs V-J diagram \citep{Labbe03}, between $z$$=1.5$ and 2.5 and model their SED using CIGALE.
The reader is referred to Ciesla et al. (2017, in prep) for a detailed description of the fitting procedure of these sources. 
Basically, we use the same procedure than in the previous Section, applying parameters that are typically used in the literature to perform a UV-to-FIR SED fitting of high redshift galaxies with CIGALE \citep[e.g.,][]{Giovannoli11,Buat15,Ciesla15}.
Without knowing the real SFH of the GOODS-South sample galaxies, we observe however several trends, similar to those we discussed in the test using the simulated galaxies.
Indeed, for the exponentially declining and delayed SFH, we observe an artificial line at high SFR resulting from the limited range of SFR probed by these two SFH, as shown in the previous section (black zone in Fig.~\ref{ms}).
An artificial line is also found at high SFR for the exponentially rising SFH, but we also note the lack of galaxies with low SFRs compared to the two other SFHs, as expected.
A different effect is found for the log-normal SFH where the range in SFR provided by the SED fitting is largely wider than when using the three other SFHs.
There is also the same perfect gradient of age on the distribution of galaxies for all of the four assumptions.
In all four models, we can see that the sources on the top of the MS seems to be the youngest at all masses and the galaxies at the bottom part of the MS the oldest ones, again regardless of the stellar mass.
We note that the age of the galaxies in the highest part of the MS is the lowest value allowed for the fit, in this case 500\,Myr, as expected from our previous figures.
This is due to the fact that, with the delayed and exponentially declining SFH, high values of SFR are only reached for small ages, as shown in Fig.~\ref{tauvage}.
However, we also note a similar gradient in the case of the exponentially rising and log-normal SFHs.
These four SFHs model a smoothly evolving main stellar population and, as shown in Sect.~\ref{outliers}, struggle to recover the true SFR of galaxies experiencing strong enhancements or decrease of their star formation activity.
The age gradient obtained with these SFH thus comes from a lack of mathematical flexibility preventing them to model recent variations of star formation activity. 
We also performed our tests using a different definition of the age, i.e. the mass-weighted age of the galaxy and retrieved the same conclusions than for the simple age definition of the SED fitting method, i.e. the presence of the gradient for every analytical forms tested here, as shown in Fig.~\ref{msz2mwa}.

\section{\label{solution}Modeling all galaxies with a single star formation history}

\begin{figure}
  	\includegraphics[width=\columnwidth]{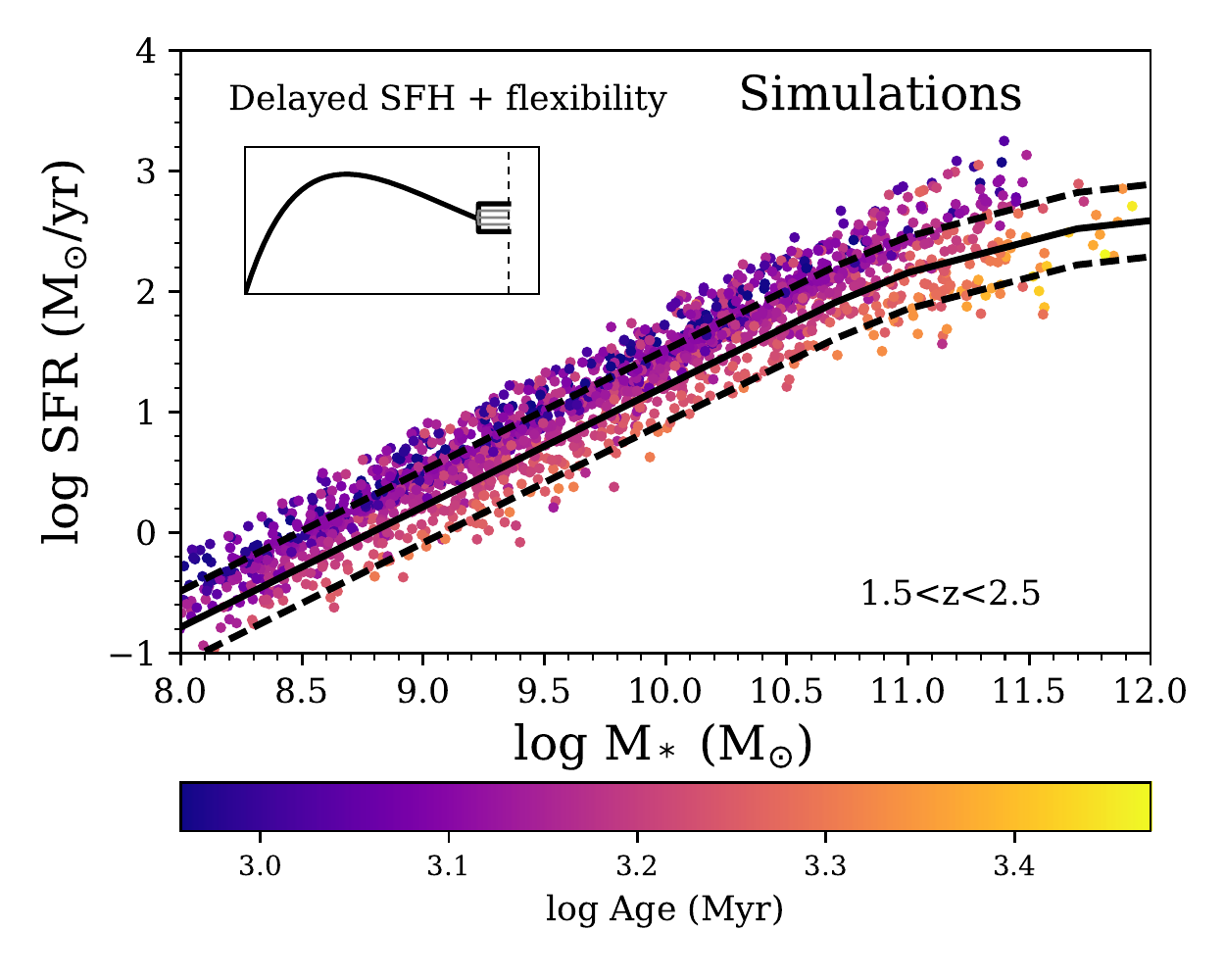}
  	\includegraphics[width=\columnwidth]{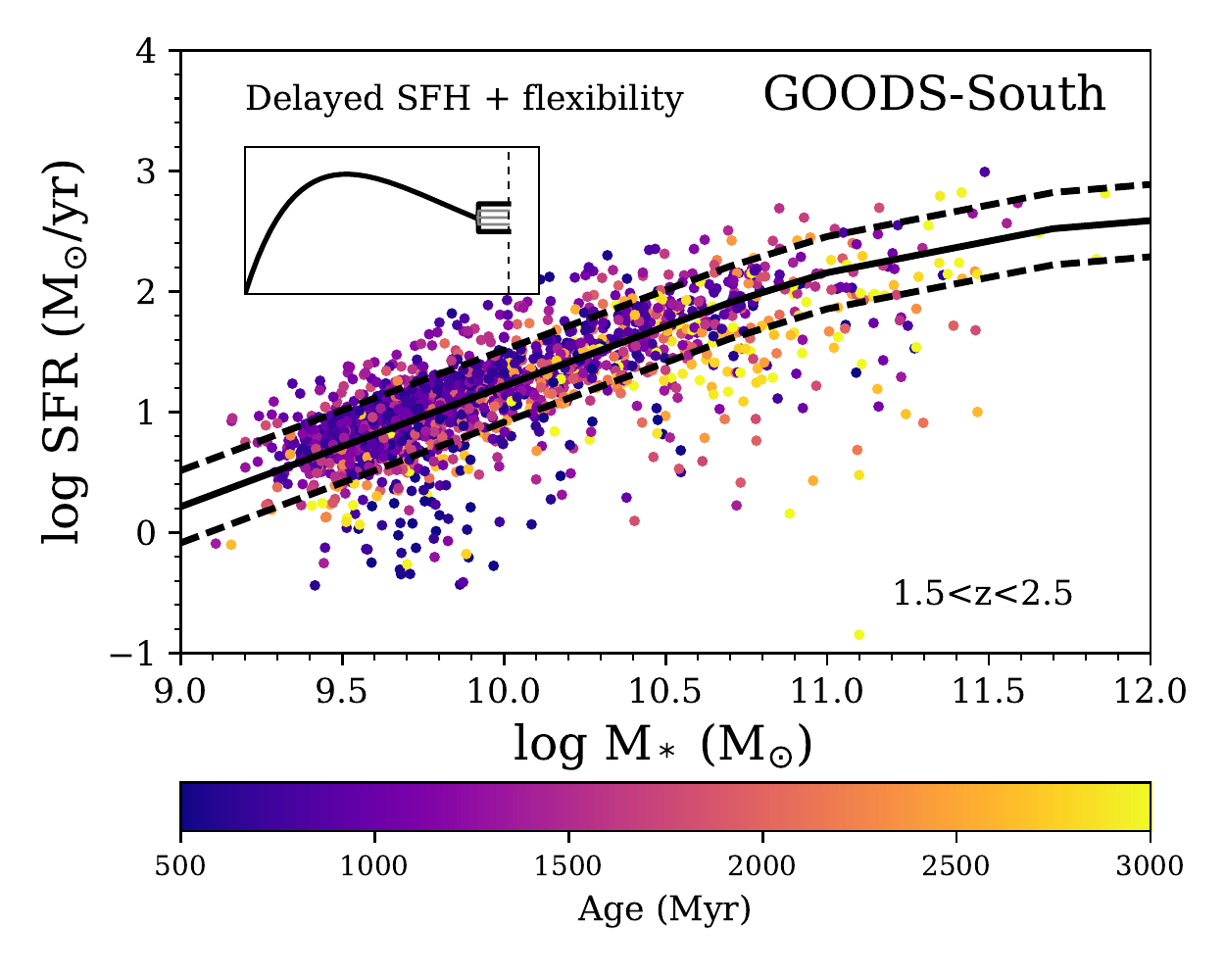}
  	\caption{ \label{msz2trunc} Top panel: SED fitting results obtained for the mock galaxy sample using the delayed+flexibility SFH, color-coded with the output age of the galaxies. Bottom panel: Main sequence relation of GOODS-South galaxies between redshift 1.5 and 2.5 obtained using the delayed SFH with an additional flexibility in the recent SFH, color-coded by the age obtained from the fitting procedure. The black line indicates the position of the MS obtained by \cite{Schreiber15} at $z$$=2$. }
\end{figure}

To recover the physical properties of galaxies, regardless of their type or star formation activity, we need to use a form that models the bulk of the star formation plus an additional flexibility in the recent SFH that can model variations in SFR within the scatter of the MS but also stronger fluctuations due to a starburst or a fast quenching event.
However, we have to keep in mind that the goal here is to derive parameters from broad-band SED fitting and thus we have to take into account that SFH parameters are not well constrained from this method.
What we propose here is to use a simple analytical form to model the bulk of the SFH and add a flexibility in the recent SFH to model small fluctuations to large variations of the SFR in the recent SFH.

\subsection{Modeling the bulk of the Star Formation History}
The ideal SFH to model the main history would be the SFH derived for the MS galaxies, since it is defined to model the bulk of galaxies.
However, as we showed in Section~\ref{ms}, such a SFH is complicated to parametrize, and the closest analytical function is handled through four free parameters.
Four parameters plus additional ones for the flexibility in the recent SFH make this mathematical form too complicated for SED modeling. 
We have to choose a simpler analytical form that can replicate the increasing part of the SFH followed by a smooth decline and provide accurate SFR measurements.
We thus exclude the exponential forms that do not model the global shape of the MS SFH and difficulties in reproducing SFR in given redshift ranges.
The form of the log-normal and delayed SFH is closest to the MS SFH, with the rising part followed by a smooth decline.
From Fig.~\ref{mscattermean}, we see that the log-normal function manages to recover the SFR of normal and star-bursting galaxies independently from redshift but fails to reproduce the SFR of rapidly quenched galaxies.
However this analytical form has two free parameters that can lead to degeneracies as we know that SFH parameters are difficult to constrain.
Indeed, we see in Fig.~\ref{msz2} that, when used on real galaxies, the log-normal SFH yields to a very dispersed MS with a large scatter.
This scatter is not compatible with what is expected from the MS computed from luminosities, infrared versus H-band, and thus independently from models assumptions \citep{Salmi12}.
To not introduce any artificial scatter in the SFR-M$_*$ relation from SED fitting, we will thus not use the log-normal SFH.
We thus propose to use a modified version of the delayed SFH that can be used at all redshifts and on all types of galaxies.
Despite limitations in recovering the SFR of very high redshift galaxies, the delayed SFH was found to be able to recover the stellar mass of galaxies, and thus to provide a good modeling of the main part of the SFH compared to exponential SFHs \citep[e.g.,][]{Lee10,Ciesla15,DaCunha15,Ciesla16}.
The rigidity implied by the mathematical shape of the SFH is problematic for the SFR but not for the stellar mass.
Indeed, testing the delayed SFH on SED built from galaxies modeled by GALFORM, \cite{Ciesla15} estimated that the stellar mass was recovered with a mean error lower than 7$\%$ and that this model better reproduced the envelope of the SFH than the exponentially decreasing models.
Furthermore, the delayed SFH also has the advantage to have only one free parameter, $\tau_{main}$.
The delayed SFH is thus suited to model the bulk of the stellar population emission.

\subsection{A flexibility in the recent SFH}
We know modify the delayed SFH in order to correct for the mathematical rigidity creating the artificial age gradient discussed in Section~\ref{bias} and allowing the SFH to reach both high and low SFR.
To overcome these limitations, we thus add a flexibility in the recent SFH to model an enhancement or decline of the SFR:
\begin{equation}	
    \mathrm{SFR}(t) \propto
    	\begin{cases}
    	  t  e^{-t/\tau_{main}}, & \text{when}\ t \leq t_0 \\
    	  \mathrm{r_{SFR}} \times \mathrm{SFR}(t=t_0), & \text{when}\ t>t_0\\
    	\end{cases}
\end{equation}	
\noindent where $t_0$ is the time when a rapid enhancement or decrease is allowed in the SFH and $r_{SFR}$ is the ratio between the SFR after $t_0$ and before $t_0$.
This flexibility allows the SFH to better probe recent variations in the SFH, regardless of its intensity (small variations within the MS scatter or intense burst or quenching).
Such a SFH was proposed in \cite{Ciesla16} and in Merlin et al. (2017, submitted), for instance, to model quenched galaxies but also in Ciesla et al. (2017, in prep) to model galaxies on the top of the MS or higher.

Values of $r_{SFR}$ larger than 1 will correspond to an enhancement of the SFR whereas values lower than 1 to a decrease.
This SFH is thus flexible enough to model galaxies inside the scatter of the MS but also starbursts and rapidly quenched galaxies \citep{Ciesla16}.
We put this SFH under the same tests than the three other SFHs except for the grid of models of Fig.~\ref{tauvage}.
Indeed, including the $r_{SFR}$ parameter all SFRs can be reached depending on the input values provided by the user.
From Fig.~\ref{mscattermean} (black filled stars), we see that this SFH recovers very well the SFR of MS galaxies, with an error lower than $\sim$10$\%$ at all redshifts.
We note that it produces SFR that are in the same order than the three other assumptions studied here at $z$$=$1 and 2, but better estimates at higher redshifts.
For starburst galaxies, the $\Delta$SFR$/$SFR values are very low, comparable to the results obtained with the rising exponential, showing a perfect recovering of starburst SFRs.
The same agreement between the true SFR and the one from the delayed SFH with flexibility is found for the quenched galaxies with an error lower than the three others at all redshifts.
We note however that there is still an error of 50$\%$ at $z$$=$1.

In Fig.~\ref{msz2trunc} (top panel), we show the results obtained with this flexible SFH on the simulated galaxies. 
There is a less pronounced artificial age gradient consistent with the actual age distribution of Fig.~\ref{mssimursb}a.
There is a gradient of age which is not parallel to the MS like what is observed for the other SFH.
The oldest galaxies are, for instance, not at the bottom of the MS, but at high masses.
This is confirmed in Fig.~\ref{mssimursb}, where we see that the gradient of age is the smallest with this flexible SFH.
However, since this simulation is an ideal case, we apply our SFH to the same SED fitting procedure applied on GOODS-South galaxies presented above (Fig.~\ref{msz2trunc}-bottom).
First, the limitation of the high values of SFR is weakened compared to other SFH assumptions and galaxies are allowed to go higher in SFR but also lower.
There is no longer a gradient of age along the MS showing that the origin of this gradient strongly depends on the assumption made on the SFH of the galaxies when performing the SED fitting.
Younger ages seem to be found at lower masses whereas older galaxies are found at higher masses.
This behavior is closer to what would be expected considering a galaxy evolving on the MS all its life with some random episodes of enhanced star formation followed by a period of lower star formation activity, as could be considered in star formation feedback scenarios, as simulated in Fig.~\ref{mssimu} (left panel).
The same result is obtained with the age defined as the mass-weighted age of the galaxy, i.e., the age gradient is no longer present.
However, since the age is known to be weakly constrained from SED fitting \citep[e.g.,][]{Buat14}, we do not interpret further this possible trend.

We conclude that adding a flexibility in the recent SFH allows for a more accurate recovering of the SFR of MS, starburst, and rapidly quenched galaxies than SFH considering one main stellar population.
The age gradient found across the MS seems to be SFH dependent and disappear with the use of this flexibility.

\section{Conclusions}
In this work, we computed the SFH of the bulk of star-forming galaxies, i.e. of galaxies following the MS all their lives.
The SFH of MS galaxies depends on cosmic time but also on the seed mass of the galaxy which can be interpreted as a proxy for the DM halo mass, itself related to its local environment.
These SFH show a peak of star formation depending on the seed mass after which the SFR smoothly declines and the stellar mass growth drastically slows down.
Following \cite{Dekel06}, these masses correspond to a hot and massive state of DM halo limiting the gas for falling on the galaxy and fueling star formation and thus possibly being at the origin of the smooth decline of the SFR.
As a consequence, we showed that MS galaxies can enter the passive region of the UVJ diagram while still forming stars following the MS.

We showed that the MS SFH is not reproduced by analytical forms usually used to perform SED fitting, neither by a log-normal SFH or a double-power-law SFH.
The best fit is provided by a Right Skew Peak Function that we use to parametrize the SFH of MS galaxies as a function of seed mass and time.

Using the SFH of MS galaxies as a benchmark, we studied the ability of exponentially rising, exponentially declining, delayed, and log-normal SFHs to retrieve the SFR of galaxies from SED fitting.
Due to mathematical limitations, the exponentially declining and delayed SFH struggle to model high SFR which starts to be problematic at $z$$>$2.
The exponentially rising and log-normal SFHs exhibit the opposite behavior with the ability to reach very high SFR but not low values such as those expected at low redshift for massive galaxies.
Simulating galaxies SED from the MS SFH using the SED modeling code CIGALE, we showed that these four analytical forms recover well the SFR of MS galaxies with an error dependent on the model and the redshift.
They are thus able to probe for small variations of SFR within the MS, such as those expected from compaction or variation of gas accretions scenarios, with an error ranging from 5 to 40$\%$ depending on the SFH assumption and the redshift.
The exponentially rising and log-normal SFHs provides very good estimates of the SFR of starburst galaxies, but all of the four assumptions fail to recover the SFR of rapidly quenched galaxies.
We note that these tests were made using information from far-IR, and using the same code to model the SED and fit them.
Although precautions were taken to minimize the potential biases, we emphasize the fact that these results should be considered as best case scenario.
Displaying the results of the SED fitting performed on the simulated galaxies on the MS diagram, we showed that all the four SFH assumptions tested in this work exhibit a gradient of age that is not an outcome of the simulation.
As the simulated galaxies are an ideal case, with only the SFH varying from one galaxy to another, we test these analytical SFHs on real data.
We use a sample of GOODS-South galaxies with redshift between 1.5 and 2.5 and show that some artificial limitations are produced at the lowest or highest SFRs depending on the model, and that the perfect gradient of age, parallel to the MS, is produced as well.

The best SFH that should be used to model galaxies is the MS SFH but, due to the complexity of its parametrization, its use for SED modeling is not reasonable as SFH parameters are unconstrained from broad band SED fitting.
Each of the four SFHs tested in this work showing some caveats in recovering high or low SFR, at different redshift, and for different galaxy populations, we propose a SFH composed of a delayed form to model the bulk of stellar population with the addition of a flexibility in the recent SFH.
This SFH provides very good estimates of the SFR of MS, starbursts, and rapidly quenched galaxies at all redshift.
Furthermore, used on the GOODS-South sample, we observe that the age gradient disappear, showing that it is dependent on the SFH assumption made to perform the SED fitting.



\begin{acknowledgements}
L.\,C. thanks C.~Schreiber, E.~Daddi, M.~Boquien, and F.~Bournaud for useful discussions.
L.\,C. warmly thanks M.~Boquien, Y.~Roehlly and D.~Burgarella for developing the new version of CIGALE on which the present work relies on.
\end{acknowledgements}

\bibliographystyle{aa}
\bibliography{letter_delayed}

\end{document}